\documentclass[preprint2]{aastex6}
\usepackage[utf8]{inputenc}
\usepackage{color}
\usepackage{enumitem}
\usepackage[fleqn]{amsmath}
\usepackage{bm}
\usepackage{url}
\usepackage{natbib}
\usepackage{csvsimple}
\usepackage{comment}
\usepackage{multirow}
\usepackage{scrextend}

\makeatletter
\def\env@matrix{\hskip -\arraycolsep 
  \let\@ifnextchar\new@ifnextchar
  \array{*{\c@MaxMatrixCols}c}}
\makeatother

\def\SPSB#1#2{\rlap{\textsuperscript{#1}}\SB{#2}}

\def\SB#1{\textsubscript{#1}}

\shorttitle{New Constraints on GJ 876}
\shortauthors{Millholland et al.}

\begin{document}

\setlength{\mathindent}{0.0pt}
\defcitealias{2010ApJ...719..890R}{R10}
\defcitealias{2010AA...511A..21C}{C10}

\title{New Constraints on Gliese 876 -- Exemplar of Mean-Motion Resonance}
\author{Sarah Millholland$^{1,6}$, Gregory Laughlin$^1$, Johanna Teske$^{2,3}$, R. Paul Butler$^2$, Jennifer Burt$^4$, Bradford Holden$^5$, Steven Vogt$^5$, Jeffrey Crane$^3$, Stephen Shectman$^3$, Ian Thompson$^3$\\}
\affil{$^1$Department of Astronomy, Yale University, New Haven, CT 06511, USA \\
$^2$ Department of Terrestrial Magnetism, Carnegie Institution for Science, Washington DC 20015, USA \\
$^3$ The Observatories of the Carnegie Institution of Washington, Pasadena, CA 91101, USA \\
$^4$ Department of Physics, Massachusetts Institute of Technology,
Cambridge, MA 02139, USA \\
$^5$ Department of Astronomy \& Astrophysics, UCO/Lick Observatory, University of California at Santa Cruz, Santa Cruz, CA 95064, USA}
\altaffiliation{$^6$ NSF Graduate Research Fellow}
\email{sarah.millholland@yale.edu}

\begin{abstract}
Gliese 876 harbors one of the most dynamically rich and well-studied exoplanetary systems. The nearby M4V dwarf hosts four known planets, the outer three of which are trapped in a Laplace mean-motion resonance. A thorough characterization of the complex resonant perturbations exhibited by the orbiting planets, and the chaotic dynamics therein, is key to a complete picture of the system's formation and evolutionary history. Here we present a reanalysis of the system using six years of new radial velocity (RV) data from four instruments.
This new data augments and more than doubles the size of the decades-long collection of existing velocity measurements. We provide updated estimates of the system parameters by employing a computationally efficient Wisdom-Holman N-body symplectic integrator, coupled with a Gaussian Process (GP) regression model to account for correlated stellar noise. 
Experiments with synthetic RV data show that the dynamical characterization of the system can differ depending on whether a white noise or correlated noise model is adopted. Despite there being a region of stability for an additional planet in the resonant chain, we find no evidence for one. Our new parameter estimates place the system even deeper into resonance than previously thought and suggest that the system might be in a low energy, quasi-regular double apsidal corotation resonance. This result and others will be used in a subsequent study on the primordial migration processes responsible for the formation of the resonant chain.  

\end{abstract}

\section{Introduction}

The nearby system of four known planets orbiting the red dwarf star Gliese 876 (GJ 876) is high on the list of landmark discoveries in the field of exoplanets. From the beginning, the discovery of the GJ 876 system brought many firsts: the first detection of a planet orbiting an M dwarf \citep{1998A&A...338L..67D, 1998ApJ...505L.147M}, the first known mean-motion resonance (MMR) in an exoplanetary system \citep{2001ApJ...556..296M}, the first super-Earth \citep{2005ApJ...634..625R}, the first resonant chain of planets \citep{2010ApJ...719..890R}, the first astrometric planetary detection \citep{2002ApJ...581L.115B}, and the first radial velocity (RV) system allowing independent determinations of the planet masses and inclinations via dynamical modeling \citep{2001ApJ...551L.109L}. The system's resonant orbital architecture permits exquisitely precise characterization rivaled only by the Solar System satellites. This makes GJ 876 a linchpin in the pursuit to understand the formation and early evolution of planetary systems.

The observational history of the GJ 876 system spans several decades and research teams. A thorough chronicle is provided by \cite{2016MNRAS.455.2484N}. Here we summarize some of the key events. The first companion, planet ``b'', a $\gtrsim 2.1 \ M_{\mathrm{Jup}}$ planet in a $\sim 61$ day orbit was discovered via Doppler velocity measurements by \cite{1998ApJ...505L.147M} and \cite{1998A&A...338L..67D}. The RV fits favored a large eccentricity for planet ``b'', but 2.5 years later \cite{2001ApJ...556..296M} showed that this apparent large eccentricity was merely disguising a second companion in a 2:1 MMR. This companion was planet ``c'', a $\gtrsim 0.56 \ M_{\mathrm{Jup}}$ planet in a $\sim 30$ day orbit.

Shortly after the discovery of planet ``c'', \cite{2001ApJ...551L.109L} and \cite{2001ApJ...558..392R} independently showed that the non-interacting Keplerian model used by \cite{2001ApJ...556..296M} was insufficient to fit the $\sim 5$ years of RV data, demonstrating that mutual perturbations between the resonant pair cause orbital evolution on rapid timescales. The apsidal lines, for example, are co-precessing at a rate of $\dot{\varpi} \sim -40^{\circ}\mathrm{yr}^{-1}$ \citep{2005AJ....129.1706F, 2005ApJ...622.1182L}. Using dynamical N-body fits, they derived more accurate constraints on the resonant pair and determined the inclination of the nearly coplanar system to be $\sim 50^{\circ}$ to the plane of the sky.

Using new Keck HIgh Resolution Echelle Spectrometer \citep[HIRES,][]{1994SPIE.2198..362V} velocities, \cite{2005ApJ...634..625R} discovered a third companion, planet ``d'', a $\sim 7.5 \ \mathrm{M_{\oplus}}$ super-Earth in a $\sim 2$ day orbit.
Several groups discussed the prospects of observing the planet(s) in transit \citep{2005ApJ...622.1182L, 2005ApJ...634..625R, 2006ApJ...653..700S, 2014ApJ...781..103K} but found no detections. 

The inclination of the system first reported by \cite{2001ApJ...551L.109L}, \cite{ 2001ApJ...558..392R} and later updated by \cite{2005ApJ...634..625R} was supported by \cite{2009AA...496..249B}, who performed a joint fit of the Keck/HIRES RVs and the astrometric data from \cite{2002ApJ...581L.115B}. 

A set of High Accuracy Radial velocity Planet Searcher (HARPS) RVs and an updated 3D orbital fit were published by \cite{2010AA...511A..21C}. Later that year, \cite{2010ApJ...719..890R} presented new Keck/HIRES RVs and announced the presence of a fourth planet, planet ``e'', a $\sim 16 \ \mathrm{M_{\oplus}}$ Neptune in a $\sim 124$ day orbit. With that finding, GJ 876 became the first known exoplanetary system to host a Laplace mean-motion resonance. \cite{2010ApJ...719..890R} found that the critical angle of the Laplace 4:2:1 resonance, $\phi_{\mathrm{Lap}} = \lambda_c - 3\lambda_b + 2\lambda_e$, librates about $0^{\circ}$ with an amplitude of $\sim 40^{\circ}$. This is distinct from the Laplace 4:2:1 resonance of the Galilean satellite system, for which the critical angle librates about $180^{\circ}$ \citep{1976ARA&A..14..215P, 1986sate.conf..159P}, and where triple conjunctions never occur.

The existence of the 4:2:1 resonant chain offers strong constraints on the system's evolutionary history. Many authors have studied a formation scenario involving convergent orbital migration through a gas disk \citep{2001A&A...374.1092S, 2002ApJ...567..596L, 2002ApJ...565..608M, 2004A&A...414..735K, 2005A&A...437..727K, 2008A&A...483..325C, 2015AJ....149..167B}. In addition to the first-order eccentricity type resonances, capture into second-order inclination resonances has also been discussed \citep{2003ApJ...597..566T, 2009ApJ...702.1662L}. Other authors have examined the resonant dynamics without reference to the system's primordial formation. Some examples include \cite{2003MNRAS.341..760B}, \cite{2003ApJ...593.1124B}, \cite{2007CeMDA..99..197V}, \cite{2013MNRAS.433..928M}.  Though the resonance is likely to be chaotic \citep{2010ApJ...719..890R,2013MNRAS.433..928M, 2015AJ....149..167B, 2016MNRAS.460.1094M}, the close encounter-preventing phase protection it provides is central in facilitating the system's long-term stability \citep{2001A&A...366..254J, 2001PASJ...53L..25K, 2002MNRAS.332..839G, 2002ApJ...572.1041J, 2003ApJ...596.1332H,  2013MNRAS.433..928M}. 

Recently, \cite{2016MNRAS.455.2484N} presented new Keck/HIRES velocities and derived updated constraints on the system using a full three-dimensional orbital model. They computed Bayes factors for model selection between three, four, and five planet models, confirming the four planets known by \cite{2010ApJ...719..890R}. They searched for a fifth planet in the system at $P \sim 15$ days but did not find strong evidence for such a planet. 

In many ways, our analysis will be similar to that of \cite{2016MNRAS.455.2484N}. The objective of our study is to offer the tightest possible parameter constraints on the system resulting from a large new RV dataset. Our primary motivation for doing so is our interest in conducting a detailed theoretical investigation into the primordial assembly of the resonant chain. As previously discussed, it is well-accepted that convergent orbital migration was required; however, not all of the system's constraints have yet been utilized to infer the particular spatial range, timescale, resonant lock order, etc. that was most likely. The variety of possible migration scenarios will be explored in a subsequent publication. First, however, we present an updated characterization of the system.

This paper is outlined as follows. In \S2, we detail the Doppler velocity dataset, which includes 332 new measurements from four spectrographs. In \S3, we outline our methodology and tools for performing fits to the RV data. This consists of a Wisdom-Holman symplectic N-body integrator, a Gaussian Process model for correlated stellar noise, and a variant of a Differential Evolution Markov Chain Monte Carlo algorithm. The results of a coplanar, four-planet fit are presented in \S4. In \S5, we present the results of a search for a $P \sim 15$ day planet, including considerations of the potential planet's long term resonant configuration and stability. In \S6, we examine the dynamical properties of the resonance and chaos suggested by our updated system characterization. We leave concluding remarks in \S7.

\section{Doppler Velocity Observations}

The dataset we consider here consists of two decades of RV measurements from four spectrographs: Keck/HIRES, HARPS, the Automated Planet Finder (APF) Levy Spectrograph, and the Carnegie Planet Finder Spectrograph (PFS) on the Magellan Telescope. Table~\ref{RV data sources} summarizes the RV observations. Note that the 54 new Keck/HIRES velocities correspond to the same observations presented by \cite{2016MNRAS.455.2484N}, but the observations presented here were reduced by the Lick-Carnegie Exoplanet Survey (LCES) team while those in \cite{2016MNRAS.455.2484N} were reduced by the California Planet Search (CPS) group. The LCES pipeline is slightly more precise than the CPS pipeline, as can be seen, for example, in the comparison of the systematic jitter estimates in Table 4 of \cite{2016MNRAS.455.2484N}. A subtable showing the first five new measurements from each instrument is shown in Table~\ref{RV data}. In addition to the RV and RV uncertainty, there is a column for the $S$-index, a measure of the emission from the Ca II H \& K spectral lines. The $S$-index measurements from each instrument have different calibration scales; this will be addressed in Section~\ref{S-index regression}. The full dataset is available in the online journal.

The Keck/HIRES, APF, and PFS RVs were obtained using the iodine cell technique and spectral synthesis procedure described by \cite{1996PASP..108..500B}. We retrieved the new HARPS spectra from the ESO-HARPS public archive and used the Template-Enhanced Radial velocity Re-analysis Application (TERRA) pipeline for the RV measurement extraction \citep{2012ApJS..200...15A}.

\begin{table}[!h]
\caption{Summary of the RV dataset. \citetalias{2010ApJ...719..890R} and \citetalias{2010AA...511A..21C} refers to \cite{2010ApJ...719..890R} and \cite{2010AA...511A..21C}, respectively.} \label{RV data sources} 
\begin{center}
\begin{tabular}{ c c c} 
 \hline
 \hline
 Instrument & Number of obs. & Date range \\
 \hline
Keck/HIRES & 168 \citepalias{2010ApJ...719..890R} + 54 (new) & 1997 - 2014 \\
HARPS & 52 \citepalias{2010AA...511A..21C}  +  204 (new) & 2003 - 2014 \\
APF & 59 (new) & 2013 - 2016 \\
PFS & 15 (new) & 2013 - 2016 \\
 \hline
\end{tabular}
\end{center}
\end{table}

\begin{table}[!h]
\caption{New Doppler velocity measurements of the GJ 876 system from four instruments. This is a sub-table, showing just five measurements per instrument. The full dataset is available in the online journal.}
\label{RV data}
\begin{center}
\begin{tabular} { c c c c }
\hline
\hline
Time [BJD] & RV [$\mathrm{m \ s^{-1}}$] & Unc [$\mathrm{m \ s^{-1}}$] & $S$-index \\
\hline
\multicolumn{4}{c}{Keck/HIRES} \\
2455339.109 & -129.04 & 1.55  & 1.08 \\
2455340.102 & -103.91 & 1.26 & 1.19 \\
2455341.114 & -47.65 & 1.86 & 1.05 \\
2455342.117 & -6.79 & 1.82 & 1.10 \\
2455370.106 & 98.92 & 1.16 & 0.96 \\
\hline
\multicolumn{4}{c}{HARPS} \\
2454770.684 & -43.04 & 0.77 & 0.98 \\
2454955.916 & -77.17 & 0.58 & 0.84 \\
2455122.628 & 65.44 & 0.82 & 0.78 \\
2455372.882 & 76.58 & 0.83 & 0.82 \\
2455373.898 & 84.85 & 0.89 & 0.84 \\
\hline
\multicolumn{4}{c}{APF} \\
2456482.935 & -178.1 & 1.95 & 0.48 \\
2456484.84 & -179.29 & 1.93 & 0.42 \\
2456486.936 & -155.44 & 1.54 & 0.46 \\
2456488.935 & -132.83 & 2.05 & 0.52 \\
2456492.951 & -87.02 & 1.69 & 0.45 \\
\hline
\multicolumn{4}{c}{PFS} \\
2456603.654 & -259.6 & 0.99 & 0.89 \\
2456605.595 & -272.58 & 1.08 & 0.89 \\
2456608.579 & -251.04 & 0.96 & 0.82 \\
2456609.602 & -245.2 & 0.92 & 0.78 \\
2456610.583 & -235.74 & 0.94 & 0.74 \\
\hline
\end{tabular}
\end{center}
\end{table}

\section{Methodology}
\subsection{Wisdom-Holman N-body symplectic integrator for RV modeling}

The authors of previous dynamical fits of the Gliese 876 system have performed the N-body calculations using a variety of schemes, including a Bulirsch-Stoer integrator \citep{2001ApJ...551L.109L, 2001ApJ...558..392R, 2005ApJ...634..625R, 2009AA...496..249B, 2010AA...511A..21C, 2010ApJ...719..890R} and a time-symmetric fourth order Hermite integrator \citep{2016MNRAS.455.2484N}. In this work, we employed a computationally efficient Wisdom-Holman symplectic integrator. Using this scheme offers a speed-up of a factor of $\sim 13$ compared to a Bulirsch-Stoer integrator using its maximum allowable timestep.

We wrote our own implementation of a second-order Wisdom-Holman integrator following the construction outlined in \cite{1991AJ....102.1528W} and \cite{1999ssd..book.....M}. We chose to do this rather than utilize publicly available code \citep[e.g.][]{2015MNRAS.452..376R} because we needed to account for the unevenly-spaced RV observations. The integrator timestep, $\Delta t_{\mathrm{WH}}$, was uniform between consecutive RV observations and no larger than 0.15 days.  In other words, the timestep was uniform between any $t_{\mathrm{obs},i}$ and $t_{\mathrm{obs},i+1}$, but variable across the time baseline. We note that this variable timestep procedure is only necessary for integrations specified to output at the RV datapoints and is not generally preferred for longer integrations. Internally, the code used units $M_{\odot}$, years, and $(4 \pi^2)^{1/3}\mathrm{AU}$ (such that $G=1$).\footnote{We defined a year to be 365.25 days, in accord with the ``Julian year'' definition, and a day to be the standard 86400 seconds. We set $G M_{\odot} = 1.3271244004193938 \times 10^{20} \ \mathrm{m^3} \mathrm{s}^{-2}$ \citep[matching the value in the REBOUND N-body software package,][]{2012A&A...537A.128R} and $G = 6.674 \times 10^{-11} \ \mathrm{m}^3 \mathrm{kg}^{-1} \mathrm{s}^{-2}$. This fixed the astronomical unit to be AU = 149595987134.062 m. It may seem unnecessary to supply such granular -- indeed millimetric -- details, but we found RV discrepancies between codes approaching $\sim$ 1 m/s when they weren't using the same set of fundamental parameters.  \nopagebreak}  

We validated our code by comparing integrations of the GJ 876 system using a Bulirsch-Stoer code \citep{1992nrfa.book.....P}, our Wisdom-Holman code, and the \texttt{WHFast} Wisdom-Holman integrator \citep{2015MNRAS.452..376R} that is part of the REBOUND N-body software package \citep{2012A&A...537A.128R}. We used \texttt{WHFast} with second-order accuracy (no symplectic correctors). The test integrations used the four-planet, coplanar best-fit parameters from \cite{2010ApJ...719..890R}\footnote{The choice of this set of parameters is representative, and the code comparison presented in this section does not depend sensitively on it.} and a timestep of 0.002 days for the Bulirsch-Stoer code. 

\begin{figure}
\epsscale{1.25}
\plotone{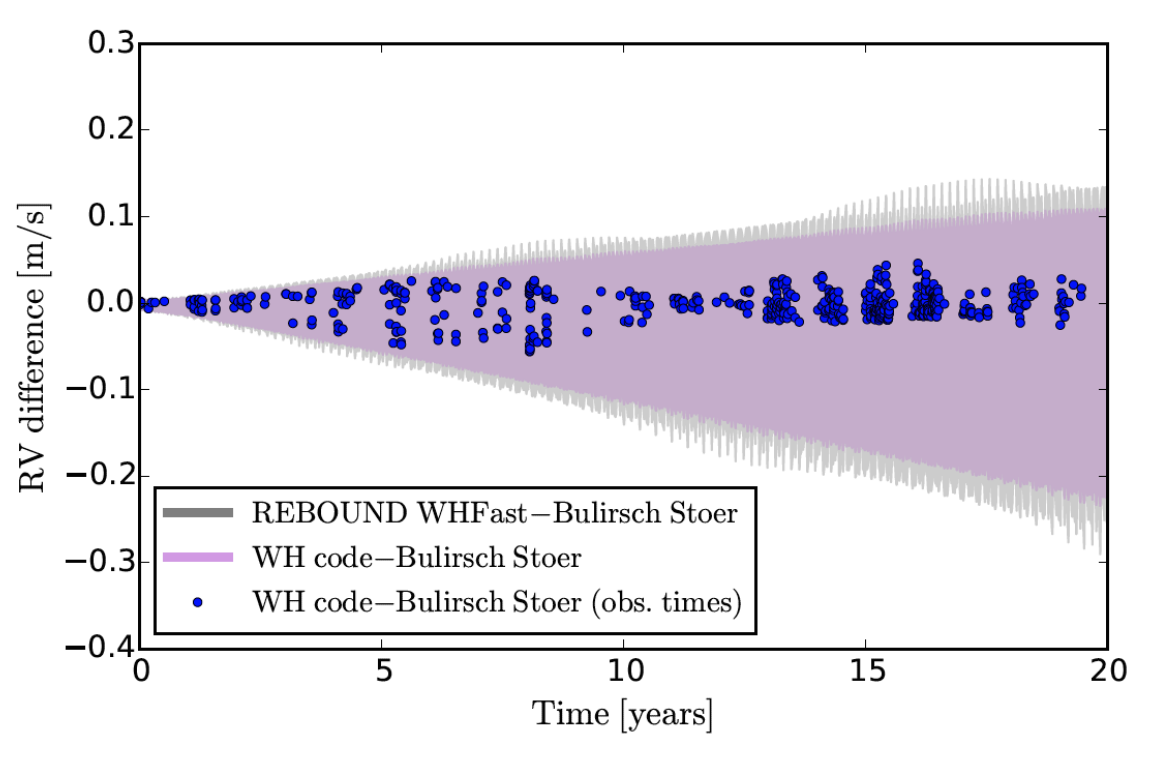}
\caption{Differences in the GJ 876 system RVs when calculated with a Bulirsch-Stoer code, REBOUND's \texttt{WHFast}, and our Wisdom-Holman code (``WH code''). For the RV differences shown by the gray and purple lines, both Wisdom-Holman codes used a uniform timestep of 0.15 days, which is about one thirteenth of the innermost planet's period. The RV differences shown by the blue dots correspond to an integration that was specified to output at the observation times of the RV dataset. In this case, the integrator timestep was uniform between consecutive RV observations and no larger than 0.15 days.} 
\label{code comparison}
\end{figure}

The comparison between the three codes is shown in Figure~\ref{code comparison}. Two types of test integrations are shown. One set of test integrations used a uniform timestep, $\Delta t_{\mathrm{WH}} =  0.15$ days, for both Wisdom-Holman codes. Their RV differences with respect to the Bulirsch-Stoer code are shown in gray and purple. The other test integration with our code was specified to output at the observation times of the RV dataset; the RV differences are shown with the blue points. 

The two uniform-timestep integrations using \texttt{WHFast} and our code perform similarly and both accumulate a phase drift relative to the Bulirsch-Stoer integration that produces absolute RV differences less than $\sim 20$ cm/s by the end of the 20 year integration.\footnote{It is worth pointing out that the magnitude of the phase error could be reduced by placing the reference epoch near the middle of the RV observational baseline and integrating both forwards and backwards in time.} The observation time integration has smaller RV differences because the timesteps were smaller than 0.15 days. Given that the RV measurement uncertainties and stellar jitter are on the order of several m/s, RV modeling discrepancies up to $\sim 20$ cm/s are acceptable, as they are a small component of the total error. We therefore used a maximum timestep of $\Delta t_{\mathrm{WH}} = 0.15$ days in the remainder of our analysis.

\subsection{Gaussian Process regression for stellar activity modeling}
\label{GP}

Most previous analyses of GJ 876 RV data -- with the exception of \cite{2011CeMDA.111..235B} -- have either not accounted for stellar activity \citep[e.g.][]{2010ApJ...719..890R}, or have added a white noise ``jitter'' term, $\sigma^2_\mathrm{jit}$, in quadrature to the reported observational uncertainties \citep{2016MNRAS.455.2484N}. This was sufficient for results derived thus far because because (1) the $\sim 0.1-5$ Gyr old \citep{2010AA...511A..21C} red dwarf is relatively quiet \citep{2005ApJ...634..625R} and (2) its rotation period \citep[$\sim 95$ days,][]{2016MNRAS.455.2484N} and harmonics are distant from the known planetary periods.

In this study, we aim to derive the most precise parameter estimates as possible and to thoroughly search for the small-amplitude signal of a potential fifth planet in the system. It is well-known that inhomogeneities on the surfaces of stars (specifically granulation, spots, plages, and faculae) produce RV signatures that can be mistakenly attributed to planetary signals. To assure that potential additional planetary signals we identify are not spurious, it is therefore worthwhile to utilize an advanced technique for modeling the correlated stellar activity-induced RVs. For this purpose we employ Gaussian Process (GP) regression.

Pioneered to the problem of radial velocity fitting by \cite{2014MNRAS.443.2517H} and \cite{2015MNRAS.452.2269R}, GP regression has become a robust and reliable technique for accounting for correlated stellar activity in RV data \citep{2015ApJ...808..127G, 2016A&A...585A.135M, 2016A&A...588A..31F, 2016A&A...593A.117A, 2016AJ....152..204L, 2017AA...599A.126D, 2017A&A...605L..11A}. GPs are used as a non-parametric method of modeling a function in some continuous input space \citep{2006gpml.book.....R, BDA}. In our case, the function is the stellar-activity RV signal. Realizations from a GP are random functions for which the function values at $n$ predetermined input values are drawn from an $n$-dimensional normal distribution with a mean vector and covariance matrix. It is common to construct the matrix from a covariance function that dictates the shrinkage towards the mean and the correlation between pairs of data points.

We employed a quasi-periodic covariance kernel, which has been used to model stellar variability in both photometric \citep[e.g.][]{2018MNRAS.474.2094A} and Doppler velocity \citep[e.g.][]{2014MNRAS.443.2517H, 2015ApJ...808..127G} applications. We denote the covariance, $k$, between observations at two different times, $t_i$ and $t_j$, by the following expression:
\begin{equation}
\begin{split}
k(t_i,t_j) &= h^2 \exp\left[-\frac{(t_i-t_j)^2}{2\lambda^2} - \frac{\sin^2\left({\pi(t_i-t_j)/\theta}\right)}{2w^2}\right]  \\
&+ ({\sigma^2_{\mathrm{obs}}}(t_i) + {\sigma^2_{\mathrm{jit, inst}}})\delta(t_i-t_j).
\end{split}
\label{covariance function}
\end{equation}

By using a quasi-periodic covariance kernel, we have assumed that the stellar activity-induced RVs have a periodic component at the stellar rotation period, $\theta$. Here $h^2$ is the amplitude of the covariance, $\lambda$ controls the coherence timescale of the variability, and $w$ dictates the relative importance of the periodic and non-periodic components. $\sigma^2_{\mathrm{obs}}(t_i)$ represents the reported observational uncertainty and $\sigma^2_{\mathrm{jit, inst}}$ is an additional white noise jitter component that is constant in time but distinct for each instrument. The Dirac delta function, $\delta(t_i-t_j)$, signifies that the second line of equation~\ref{covariance function} only applies to the main diagonal of the covariance matrix. 

Let $\mathbf{K}$ be the covariance matrix constructed from the covariance function (i.e., $K_{ij} = k(t_i,t_j)$) with the set of covariance hyperparameters $\bm{\phi} = (h^2, \lambda, \theta, w, \sigma^2_{\mathrm{jit, inst_1}}, ... , \sigma^2_{\mathrm{jit, inst_5}}).$ In addition, let $\mathbf{r}$ be the vector of RV residuals obtained after subtracting both the instrument-specific zero-point offset, $\gamma_{\mathrm{inst}}$, and the N-body model velocities calculated with the vector, $\bm{\theta}$, of system parameters (periods, eccentricities, etc.). Explicitly, the components of $\mathbf{r}$ are
\begin{equation}
\label{RV residuals equation}
\mathrm{r}_i = \mathrm{RV}_{\mathrm{obs}}(t_i) - \gamma_{\mathrm{inst}} - \mathrm{RV}_{\mathrm{N-body}}(t_i;\bm{\theta}).
\end{equation}
The residuals $\mathbf{r}$ thus capture the stellar-activity induced RVs and measurement noise.

With this notation in hand, the log-likelihood may be written as
\begin{equation}
\ln \mathcal{L}(\mathbf{r}|\bm{\theta}, \bm{\gamma}, \bm{\phi}) = -\frac{1}{2}\mathbf{r}^{T}\mathbf{K}^{-1}\mathbf{r}-\frac{1}{2}\ln(\det \mathbf{K}) -\frac{n}{2}\ln(2\pi).
\end{equation}
Given specifications for the prior distributions, the posterior distribution of model parameters may be determined using, for example, Markov Chain Monte Carlo (MCMC) sampling. This is discussed in the subsequent section.

We performed all calculations associated with the GP regression with the publicly available \textit{george}\footnote{\href{http://dan.iel.fm/george/current/}{http://dan.iel.fm/george/current/}} code \citep{2015ITPAM..38..252A}.

Before proceeding, we briefly note that the technique used by \cite{2011CeMDA.111..235B} for accounting for red noise in GJ 876 data was essentially a GP model with covariance function (here using different notation),
\begin{equation}
\begin{split}
k(t_i,t_j) &= h^2 \exp\left(\frac{-|t_i-t_j|}{\lambda}\right) + \\  &({\sigma^2_{\mathrm{obs}}}(t_i) + {\sigma^2_{\mathrm{jit, inst}}})\delta(t_i-t_j).
\end{split}
\end{equation}
However, the language associated with the GP formulation was not yet popular in RV analyses at that time.

\subsection{Differential Evolution Markov Chain Monte Carlo for parameter estimation}
\label{DE-MCMC}

With the N-body dynamical modeling and stellar activity modeling in place, we now describe our adopted methodology for posterior parameter estimation. We used a Differential Evolution Markov Chain Monte Carlo (DE-MCMC) method \citep{TerBraak2006}. DE-MCMC is an ensemble sampler that iterates many Markov Chains in parallel and uses the inter-chain differences to inform the parameter jumps from one iteration to the next. Like \cite{2016MNRAS.455.2484N}, we used $N = 300$ chains and an adaptive jumping scale parameter (the parameter called $\gamma$ in \citealt{TerBraak2006}). The new proposal state of chain $i$ is generated using the displacement vector between the states of two randomly selected chains:
\begin{equation}
X_{i, \mathrm{proposal}} = X_i + \gamma(X_{r_1} - X_{r_2}).
\end{equation}

The scale parameter is given by $\gamma = \gamma_0 (1 + z)$ with $z$ a Gaussian random variable with standard deviation $\sigma_{\gamma}$. We fixed $\sigma_{\gamma} = 0.05$. Following \cite{TerBraak2006}, we initially set $\gamma_0 = 2.38/\sqrt{2 n_{\mathrm{dim}}}$. We then adjusted $\gamma_0$ each generation to maintain the acceptance fraction in the range 10-30\%; if the acceptance fraction in a given generation was too high/low, $\gamma_0$ was scaled up/down by 5\%\footnote{As a matter of practical importance for those interested in adopting similar techniques, we note that the adaptive scaling results in $\gamma_0$ being quite small during burn-in ($\sim 0.02$). It then increases by about an order of magnitude as burn-in finishes and the chains converge. }. Every tenth generation, we set $\gamma = 1.0$ in order to allow the chains to jump between different modes in the case of multimodal posterior distributions \citep{TerBraak2006, 2014ApJS..210...11N}.

During the burn-in period, we also applied an outlier rejection procedure inspired by that of the popular DiffeRential Evolution Adaptive Metropolis (DREAM) algorithm \citep{VrugtEtAl:2009}. This extra step rapidly decreased convergence times. After each generation during burn-in, we calculated the upper and lower quartiles of the distribution of chain log-likelihoods and considered outliers to be those with $\log \mathcal{L} < \mathrm{Q_1} - 2(\mathrm{IQR})$, where IQR is the interquartile range. The outliers were then replaced with a perturbation of three randomly selected remaining chains: $X_{\mathrm{new}} = X_{r_1} + \gamma(X_{r_2} - X_{r_3})$, where $r_1$, $r_2$, and $r_3$ represent random, distinct chain indices. This outlier rejection procedure does not retain detailed balance, which is why it can only be applied during burn-in.

We parallelized the code by first randomly splitting the chains into two sets of equal size; we advanced in parallel all chains in one set while using the positions of the chains in the \textit{opposite} set for choosing the jumping vectors. This is identical to the parallelization procedure implemented for the affine-invariant ensemble MCMC sampler by \cite{2013PASP..125..306F}.

The parameters involved in the RV fits are the planet periods, RV semi-amplitudes, eccentricities, arguments of periastron, mean anomalies, and inclinations, the instrumental jitter parameters and zero-point offsets, and the GP covariance hyperparameters (see Section~\ref{GP}). The following set of transformed parameters were introduced to reduce correlations between parameters:
$\sqrt{e}\cos{\omega}$, $\sqrt{e}\sin{\omega}$, $\sqrt{K}\cos{(\omega + M)}$, $\sqrt{K}\sin{(\omega + M)}$. All other parameters were left untransformed. The Jacobian of the transformation is unity and therefore may be ignored. The initial and prior distributions will be discussed in Section~\ref{four-planet fit}.

\subsection{GP regression of the $S$-index time series}
\label{S-index regression}

We are now fully equipped to analyze the RV data using the N-body integrator, stellar activity model, and DE-MCMC sampler outlined in the previous sections. Before proceeding, however, we first analyze a set of ancillary stellar data to gain initial information about the properties of the stellar activity. This information will be used to place priors on the GP hyperparameters. Previous studies have considered photometry \citep[e.g.][]{2014MNRAS.443.2517H, 2015ApJ...808..127G} or spectroscopic measures \citep[e.g.][]{2016A&A...593A.117A} as constraints on the activity for the GP modeling. These ancillary data can be fit jointly or separately from the RVs; here we choose to fit them separately. 

\begin{figure*}[!t]
\epsscale{1.2}
\plotone{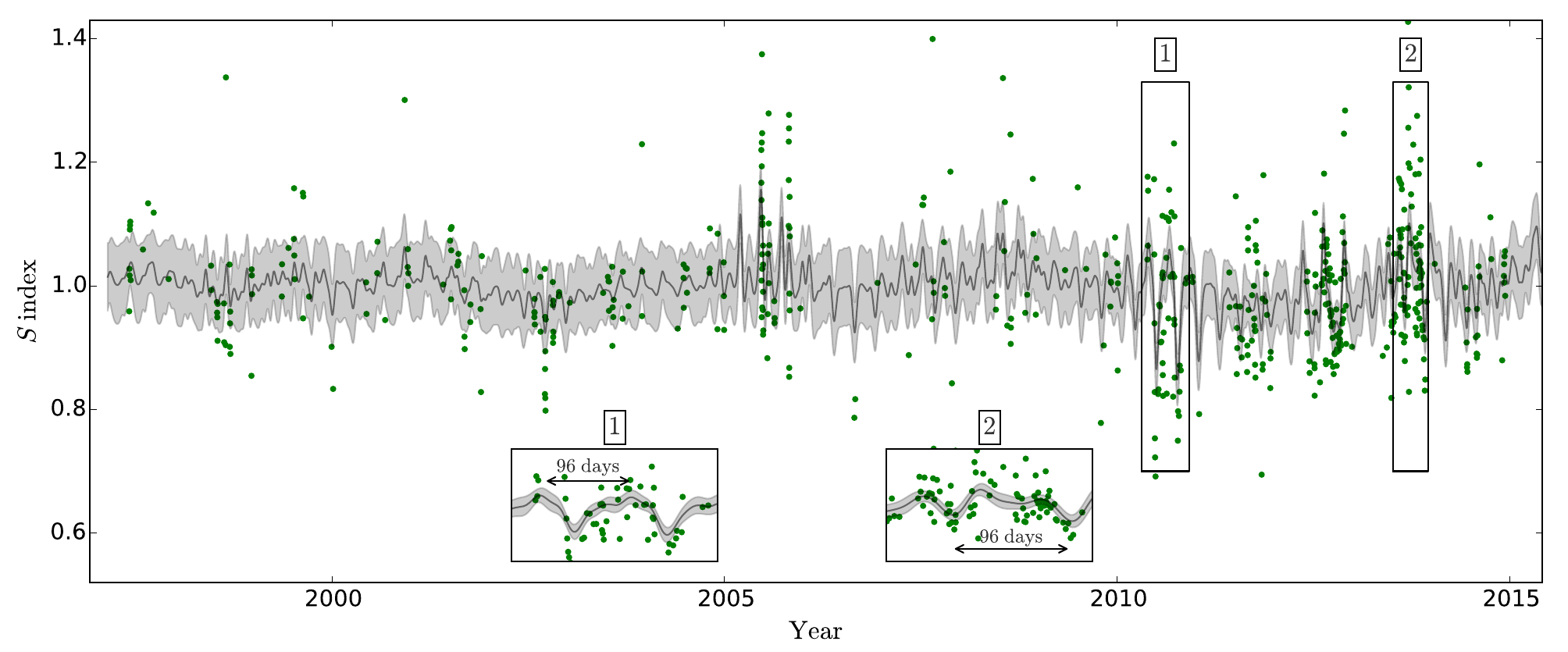}
\caption{$S$-index observations versus time, as measured by the Keck/HIRES, HARPS, APF, and PFS spectrographs. In the Gaussian Process regression, the measurements were fit to a zero mean, but this plot shows the measurements with a vertical offset equal to one. The gray line shows the regression results, specifically the mean of the posterior predictive distribution conditioned on the data. The gray band is plus/minus the square root of the diagonal of the predictive covariance matrix. The inset figures show zoomed-in views of two sections of data. In these insets, we labeled the $96.6 \pm \SPSB{3.6}{3.7} \ \mathrm{days}$ stellar rotation period estimate obtained from the $S$-index fit.} 
\label{S-index fit}
\end{figure*}

The first piece of ancillary information is the stellar rotation period, which has been estimated in previous studies. \cite{2005ApJ...634..625R} used high-precision photometric modeling and measured it to be $\sim 97$ days. \cite{2016MNRAS.455.2484N} used HARPS spectra measurements of the activity-sensitive H$\alpha$ absorption line and measured the periodicity to be $\sim 95$ days. These prior estimates of the rotation period can be used as an informative constraint within the GP regression.

In addition to the rotation period, we use observations of the $S$-index, a measure of the emission from the Ca II H \& K spectral lines and a well-known proxy for chromospheric magnetic activity. The $S$-index measurements from the four spectrographs are not expected to be normalized uniformly due to different calibration procedures. To account for this, we allowed the scale and offset of observations from each instrument to be free parameters in the fit. The transformed observation from a given instrument, $s_{\mathrm{trans}, i}$, was therefore computed as $s_{\mathrm{trans}, i} = m_{\mathrm{inst}}s_{\mathrm{raw}, i} + b_{\mathrm{inst}}$ with $m_{\mathrm{inst}}$ and $b_{\mathrm{inst}}$ free parameters. An exception was the KECK/HIRES data, which had $m_{\mathrm{HIRES}}$ fixed to equal 1. In effect, the observations from HARPS, APF, and PFS were calibrated to the scale of the Keck/HIRES data.

We applied a GP regression with a quasi-periodic covariance kernel, where now the log-likelihood is given by 
\begin{equation}
\ln \mathcal{L}(\mathbf{s}|\bm{\phi}) = -\frac{1}{2}\mathbf{s}^{T}\mathbf{K}^{-1}\mathbf{s}-\frac{1}{2}\ln(\det \mathbf{K}) -\frac{n}{2}\ln(2\pi).
\end{equation}
Here $\mathbf{s}$ is the vector of the scaled and offset $S$-index measurements, where the instrumental offsets will converge to values that make the mean of $\mathbf{s}$ equal to zero. The covariance function (see equation~\ref{covariance function}) did not include observational uncertainties (i.e. no $\sigma^2_{\mathrm{obs}}(t_i)$ term), but it included a constant white-noise jitter parameter that was not instrument-specific. The free parameters were the GP covariance parameters and the instrument-dependent scale and offset parameters, $m_{\mathrm{inst}}$ and $b_{\mathrm{inst}}$. 

We used DE-MCMC for parameter estimation with $N=50$ chains. The prior distributions and parameter estimates resulting from the MCMC sampling are shown in Table~\ref{S-index fit parameters}. Of particular interest, we note that the fit identifies a $96.6 \pm \SPSB{3.6}{3.7} \ \mathrm{days}$ stellar rotation period, which is consistent with previous estimates. Figure~\ref{S-index fit} shows the regression results with the best-fit kernel parameters.  The results for the parameters $\lambda$, $\theta$, and $w$ will be used as priors in the RV fitting. The $h^2$ and $\sigma^2_{\mathrm{jit}}$ estimates will not used, however, because they are connected to the scale of the $S$-index measurements, which is distinct from the scale of the RV measurements. 

\begin{table}[!h]
\caption{Prior distributions and parameter estimates of the GP regression to the $S$-index time series. The estimates reported are the posterior distribution means, and the lower and upper uncertainties are the 16\textsuperscript{th} and 84\textsuperscript{th} percentiles.} 
\label{S-index fit parameters} 
\begin{center}
\begin{tabular}{ c  c  c} 
 \hline
 \hline
\multicolumn{3}{c}{GP covariance parameters} \\
\hline
 Parameter & Prior distribution & Estimate \\
 \hline
$h$ & $h \sim \mathcal{U}[0, 0.25]$ & 
$0.063\pm\SPSB{0.008}{0.008}$  \\ 
$\lambda \ \mathrm{[days]}$ & $\lambda \sim \mathcal{U}[1, 600]$ & $125\pm\SPSB{50}{50}$ \\ 
$\theta \ \mathrm{[days]}$ & $\ln \theta \sim \mathcal{U}[\ln75, \ln120]$ & $(\theta = \ ) 96.6\pm\SPSB{3.6}{3.7}$ \\ 
$w$ & $w \sim \mathcal{U}[0.05, 10]$ & $0.29\pm\SPSB{0.10}{0.12}$ \\ 
$\sigma_{\mathrm{jit}}$ & $\sigma_{\mathrm{jit}} \sim \mathcal{U}[0, 0.2]$ & $0.11\pm\SPSB{0.004}{0.004}$ \\ 
\hline
\hline
\multicolumn{3}{c}{Scale and offset parameters} \\
\hline
 Parameter & Prior distribution & Estimate \\
\hline
$m_{\mathrm{HIRES}}$ & -- & 1.0 (fixed) \\
$b_{\mathrm{HIRES}}$ & $b_{\mathrm{HIRES}} \sim \mathcal{U}[-2,0]$ & $-1.02\pm\SPSB{0.016}{0.014}$ \\
$m_{\mathrm{HARPS}}$ & $m_{\mathrm{HARPS}} \sim \mathcal{U}[1, 3]$ & $1.004\pm\SPSB{0.003}{0.003}$ \\
$b_{\mathrm{HARPS}}$ & $b_{\mathrm{HARPS}} \sim \mathcal{U}[-2,0]$ & $-0.83\pm\SPSB{0.01}{0.009}$\\
$m_{\mathrm{APF}}$ & $m_{\mathrm{APF}} \sim \mathcal{U}[1, 3]$ & $2.14\pm\SPSB{0.10}{0.11}$ \\
$b_{\mathrm{APF}}$ & $b_{\mathrm{APF}} \sim \mathcal{U}[-2,0]$ & $-0.95\pm\SPSB{0.049}{0.048}$ \\
$m_{\mathrm{PFS}}$ & $m_{\mathrm{PFS}} \sim \mathcal{U}[1, 3]$ & $1.07\pm\SPSB{0.052}{0.056}$ \\
$b_{\mathrm{PFS}}$ & $b_{\mathrm{PFS}} \sim \mathcal{U}[-2,0]$ & $-0.87\pm\SPSB{0.063}{0.063}$ \\
\end{tabular}
\end{center}
\end{table}

\section{four-planet fit}
\label{four-planet fit}

We first report the results of a coplanar, four-planet fit to the data. In this paper, we only consider coplanar configurations. \cite{2016MNRAS.455.2484N} performed several different three-dimensional fits, including those that enforced dynamical stability for up to $10^7$ years. Their results strongly favor a very flat system configuration ($|{i_c - i_b}| < 2.6^{\circ}$, $|{i_b - i_e}| < 7.8^{\circ}$) and suggest that long-term orbital stability is very sensitive to the planets' mutual inclinations. (See in particular their Figure 4.)

\begin{figure*}[!t]
\epsscale{1.25}
\plotone{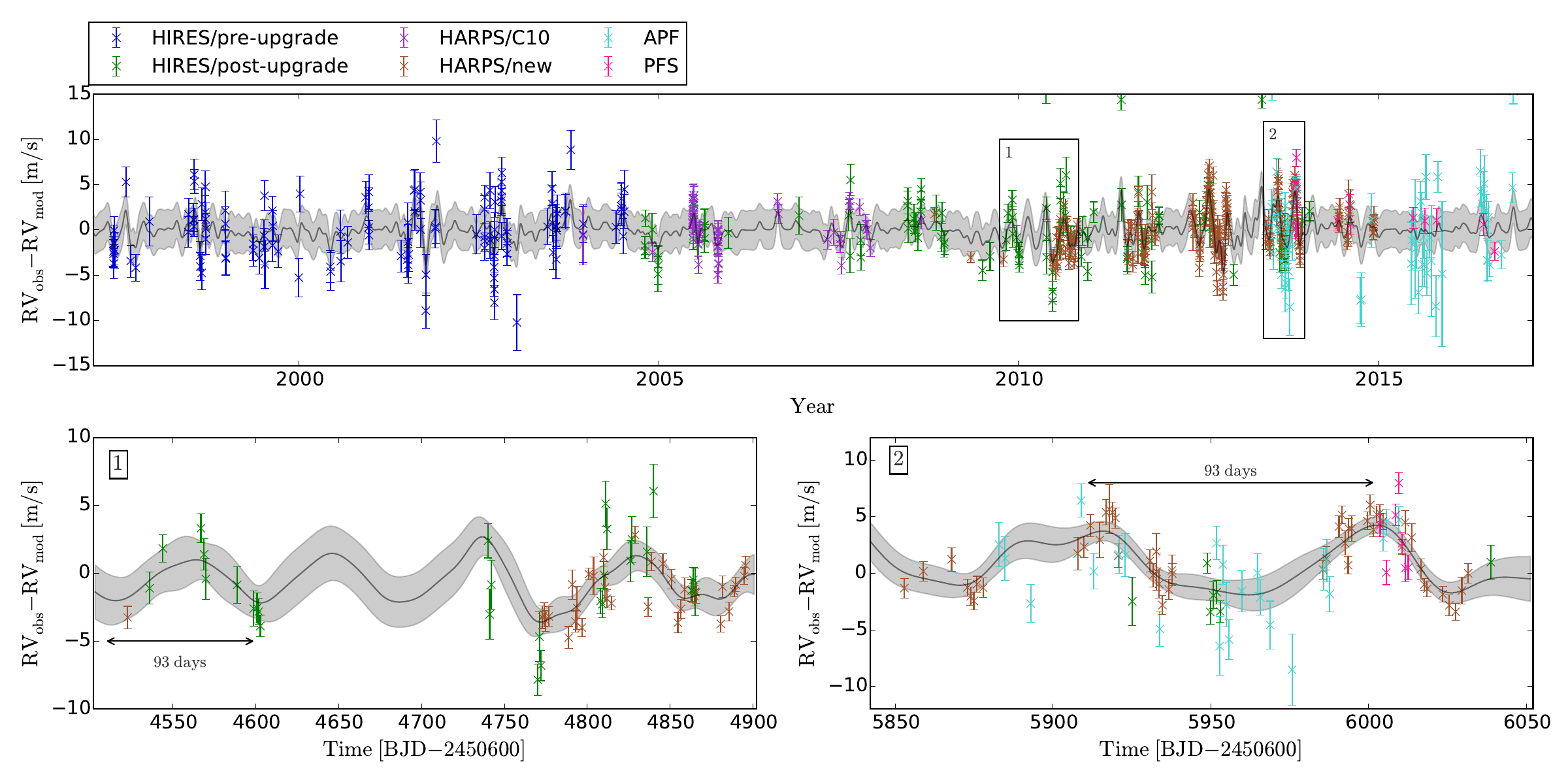}
\caption{The RV residuals, $\mathrm{RV_{obs}} - \mathrm{RV_{mod}}$, obtained by subtracting the instrumental zero-point offsets and best-fit dynamical N-body model from the data. The remaining signal corresponds to the stellar activity-induced radial velocities and measurement noise. The solid gray line is the mean of the posterior predictive distribution conditioned on the residuals, and the shaded region represents plus/minus the square root of the diagonal of the predictive covariance matrix. The two zoomed-in panels in the lower half of the figure highlight the quasi-periodic variability at the $92.6 \pm \SPSB{3.3}{3.2} \ \mathrm{days}$ days stellar rotation period estimated from the RV fit.} 
\label{RV residuals}
\end{figure*}

In addition to providing support for our assumption of a coplanar configuration, the results of previous studies of the system allow us to concentrate on a relatively small MCMC sampling domain. We initialized the DE-MCMC Markov Chains using Gaussian distributions with means equal to the best-fit coplanar parameters from \cite{2010ApJ...719..890R} and standard deviations equal to three times the reported uncertainties of  \cite{2016MNRAS.455.2484N}'s coplanar, four-planet fit\footnote{This may seem like an unusual decision, but we are using the same epoch as \cite{2010ApJ...719..890R}, so it makes sense to use their parameter estimates for the means. The choice is not crucial because all parameters converge well within the initial distributions.}. We used uniform priors on all variables and imposed lower and upper physical boundaries where appropriate (e.g. orbital periods no smaller than zero). In addition, the results of Section~\ref{S-index regression} were used to inform the boundaries of the GP covariance parameters; the priors for these parameters were $h [\mathrm{m/s}] \sim \mathcal{U}[0.5, 10]$, $w \sim \mathcal{U}[0.1, 10]$, $\ln(\theta [\mathrm{days}]) \sim \mathcal{U}[\ln85, \ln100]$, and $\lambda [\mathrm{days}] \sim \mathcal{U}[10, 300]$. To test the chains for non-convergence, we used the Gelman-Rubin $\hat{R}$ convergence diagnostic \citep{1992StaSc...7..457G}. We considered the burn-in period to be complete once the mean $\hat{R}$ value over all dimensions was below 1.1.

The parameter estimates resulting from the coplanar, four-planet fit are displayed in Table~\ref{four-planet fit parameters}. A down-sampled subset of the posterior samples is available at \href{https://github.com/smillholland/GJ876}{https://github.com/smillholland/GJ876} and archived at \href{https://doi.org/10.5281/zenodo.1149601}{10.5281/zenodo.1149601}. We find parameter estimates generally consistent with \cite{2016MNRAS.455.2484N} but with uncertainties a factor of 1.6 smaller on average. 

The GP covariance parameters, $\theta$, $\lambda$ and $w$, are consistent with those of the $S$-index fit within uncertainties. In the estimation of these parameters, we ignored samples in a disfavored local minimum with $\lambda < 20$ days and/or with $w > 0.7$. This was based on our assumption that the RV noise should be correlated and therefore should not have small $\lambda$; the results may be sensitive to this assumption. The consistency between the GP parameter estimates in the $S$-index and RV fits provides confidence that we are indeed extracting stellar activity-induced RV noise. The comparison of the $\sim93$-day rotation period with the $\sim72$-day coherence timescale suggests that active regions on GJ 876 evolve on somewhat shorter timescales than a single rotation period. We also remark that we did not expect to see $\sigma_{\mathrm{jit, HIRES_{pre}}} < \sigma_{\mathrm{jit, HIRES_{post}}}$. This might be related to the handful of HIRES/post-upgrade outliers that are shown at the upper edge of the plot in Figure~\ref{RV residuals}.

One planet parameter estimate of particular interest is the eccentricity of planet ``e'' ($0.057 \pm 0.039$), which is noticeably smaller than $0.119 \pm 0.05$ from \cite{2016MNRAS.455.2484N} or $0.207 \pm 0.055$ from \cite{2010ApJ...719..890R}. (Note that our epoch is the same as \citealt{2010ApJ...719..890R} but different from \citealt{2016MNRAS.455.2484N}.) This reduction in the estimate of $e_d$ may be related to our adoption of the GP model for correlated noise \citep{2011CeMDA.111..235B}. Interestingly, our fit still suggests a non-zero value of $e_d$, but it is not as extreme as previously determined. This observation has relevance for understanding the tidal interactions between planet ``d'' and the host star, and their relation to eccentricity pumping induced by secular interactions with the resonant outer planets.

Figure~\ref{RV residuals} displays the residuals of the fit. The residuals account for the instrumental zero-point offsets and the dynamical N-body fit (see equation~\ref{RV residuals equation}), such that the remaining signal is the stellar activity-induced RVs and measurement noise. The superimposed mean of the posterior predictive distribution displays the quasi-variability at the stellar rotation period, along with a smaller, longer-timescale variability. The comparison of the $S$-index fit (Figure~\ref{S-index fit}) to the RV fit (Figure~\ref{RV residuals}) demonstrates again that we recover stellar activity noise at a rotation period that is consistent with previous estimates \citep{2005ApJ...634..625R, 2016MNRAS.455.2484N}. The figure also illustrates the particularly high quality of the HARPS and PFS velocities.

\setlength{\extrarowheight}{5pt}
\begin{table*}[!t]
\label{four-planet fit parameters} 
\begin{center}
\begin{longtable}{ c | c  c  c  c } 
 \hline
 \hline
 \multicolumn{5}{c}{Planet parameters} \\
 \hline
 Parameter & Planet $d$ & Planet $c$ & Planet $b$ & Planet $e$ \\
 \hline
 \begin{filecontents*}{means_and_errorbars_112517.csv}
POneMean, POneStd, POneLower, POneUpper, PTwoMean, PTwoStd, PTwoLower, PTwoUpper, PThreeMean, PThreeStd, PThreeLower, PThreeUpper, PFourMean, PFourStd, PFourLower, PFourUpper, aOneMean, aOneStd, aOneLower, aOneUpper, aTwoMean, aTwoStd, aTwoLower, aTwoUpper, aThreeMean, aThreeStd, aThreeLower, aThreeUpper, aFourMean, aFourStd, aFourLower, aFourUpper, KOneMean, KOneStd, KOneLower, KOneUpper, KTwoMean, KTwoStd, KTwoLower, KTwoUpper, KThreeMean, KThreeStd, KThreeLower, KThreeUpper, KFourMean, KFourStd, KFourLower, KFourUpper, mOneMean, mOneStd, mOneLower, mOneUpper, mTwoMean, mTwoStd, mTwoLower, mTwoUpper, mThreeMean, mThreeStd, mThreeLower, mThreeUpper, mFourMean, mFourStd, mFourLower, mFourUpper, eOneMean, eOneStd, eOneLower, eOneUpper, eTwoMean, eTwoStd, eTwoLower, eTwoUpper, eThreeMean, eThreeStd, eThreeLower, eThreeUpper, eFourMean, eFourStd, eFourLower, eFourUpper, argperOneMean, argperOneStd, argperOneLower, argperOneUpper, argperTwoMean, argperTwoStd, argperTwoLower, argperTwoUpper, argperThreeMean, argperThreeStd, argperThreeLower, argperThreeUpper, argperFourMean, argperFourStd, argperFourLower, argperFourUpper, MOneMean, MOneStd, MOneLower, MOneUpper, MTwoMean, MTwoStd, MTwoLower, MTwoUpper, MThreeMean, MThreeStd, MThreeLower, MThreeUpper, MFourMean, MFourStd, MFourLower, MFourUpper, ArgperPlusMOneMean, ArgperPlusMOneStd, ArgperPlusMOneLower, ArgperPlusMOneUpper, ArgperPlusMTwoMean, ArgperPlusMTwoStd, ArgperPlusMTwoLower, ArgperPlusMTwoUpper, ArgperPlusMThreeMean, ArgperPlusMThreeStd, ArgperPlusMThreeLower, ArgperPlusMThreeUpper, ArgperPlusMFourMean, ArgperPlusMFourStd, ArgperPlusMFourLower, ArgperPlusMFourUpper, ecosomegaOneMean, ecosomegaOneStd, ecosomegaOneLower, ecosomegaOneUpper, ecosomegaTwoMean, ecosomegaTwoStd, ecosomegaTwoLower, ecosomegaTwoUpper, ecosomegaThreeMean, ecosomegaThreeStd, ecosomegaThreeLower, ecosomegaThreeUpper, ecosomegaFourMean, ecosomegaFourStd, ecosomegaFourLower, ecosomegaFourUpper, esinomegaOneMean, esinomegaOneStd, esinomegaOneLower, esinomegaOneUpper, esinomegaTwoMean, esinomegaTwoStd, esinomegaTwoLower, esinomegaTwoUpper, esinomegaThreeMean, esinomegaThreeStd, esinomegaThreeLower, esinomegaThreeUpper, esinomegaFourMean, esinomegaFourStd, esinomegaFourLower, esinomegaFourUpper, nodeOneMean, nodeOneStd, nodeOneLower, nodeOneUpper, nodeTwoMean, nodeTwoStd, nodeTwoLower, nodeTwoUpper, nodeThreeMean, nodeThreeStd, nodeThreeLower, nodeThreeUpper, nodeFourMean, nodeFourStd, nodeFourLower, nodeFourUpper, iOneMean, iOneStd, iOneLower, iOneUpper, iTwoMean, iTwoStd, iTwoLower, iTwoUpper, iThreeMean, iThreeStd, iThreeLower, iThreeUpper, iFourMean, iFourStd, iFourLower, iFourUpper, offsetOneMean, offsetOneStd, offsetOneLower, offsetOneUpper, offsetTwoMean, offsetTwoStd, offsetTwoLower, offsetTwoUpper, offsetThreeMean, offsetThreeStd, offsetThreeLower, offsetThreeUpper, offsetFourMean, offsetFourStd, offsetFourLower, offsetFourUpper, offsetFiveMean, offsetFiveStd, offsetFiveLower, offsetFiveUpper, offsetSixMean, offsetSixStd, offsetSixLower, offsetSixUpper, SigmaJitOneMean, SigmaJitOneStd, SigmaJitOneLower, SigmaJitOneUpper, SigmaJitTwoMean, SigmaJitTwoStd, SigmaJitTwoLower, SigmaJitTwoUpper, SigmaJitThreeMean, SigmaJitThreeStd, SigmaJitThreeLower, SigmaJitThreeUpper, SigmaJitFourMean, SigmaJitFourStd, SigmaJitFourLower, SigmaJitFourUpper, SigmaJitFiveMean, SigmaJitFiveStd, SigmaJitFiveLower, SigmaJitFiveUpper, SigmaJitSixMean, SigmaJitSixStd, SigmaJitSixLower, SigmaJitSixUpper, hMean, hStd, hLower, hUpper, wMean, wStd, wLower, wUpper, logProtMean, logProtStd, logProtLower, logProtUpper, LambdaMean, LambdaStd, LambdaLower, LambdaUpper
1.937793,0.000008,0.000008,0.000008,30.0972,0.0074,0.0073,0.0071,61.1057,0.0077,0.0074,0.0074,123.83,0.73,0.66,0.7,0.021838,0.000000064,0.000000065,0.000000063,0.136044,0.000022,0.000022,0.000021,0.218627,0.000018,0.000017,0.000017,0.3501,0.0014,0.0012,0.0013,6.02,0.17,0.17,0.17,87.46,0.28,0.28,0.28,211.57,0.3,0.29,0.3,3.13,0.35,0.34,0.34,7.55,0.23,0.23,0.23,265.6,2.7,2.7,2.7,845.2,9.7,9.4,9.5,15.8,1.8,1.7,1.7,0.057,0.037,0.039,0.039,0.2571,0.0019,0.0019,0.0019,0.0325,0.0018,0.0017,0.0016,0.03,0.022,0.024,0.023,116,57.7,46.7,47.3,51.09,0.82,0.78,0.77,35.5,4.2,4.4,4.1,164.1,74.1,51.4,90.1,104,57.3,46.9,46.7,292.55,1.01,0.99,1,340.6,4.2,4,4.3,50.3,75.3,86.8,46,220.7,3.7,3.8,3.7,343.64,0.44,0.44,0.43,16.06,0.24,0.23,0.23,208.8,7.1,6.3,6.4,-0.019,0.033,0.033,0.029,0.1615,0.0034,0.0032,0.0033,0.0264,0.0023,0.0019,0.0019,-0.016,0.024,0.025,0.019,0.041,0.039,0.04,0.041,0.2,0.0023,0.0023,0.0023,0.0188,0.002,0.002,0.002,0.004,0.023,0.014,0.019,0,0,0,0,0,0,0,0,0,0,0,0,0,0,0,0,53.06,0.88,0.85,0.85,36.94,0.88,0.85,0.85,36.94,0.88,0.85,0.85,36.94,0.88,0.85,0.85,94.58,0.66,0.66,0.67,-4.6,0.44,0.44,0.44,25.39,0.5,0.49,0.48,28.75,0.47,0.47,0.48,81.13,0.81,0.8,0.78,-20.88,0.96,0.95,0.95,1.13,0.23,0.22,0.22,2.33,0.15,0.15,0.15,1.87,0.32,0.31,0.32,2.84,0.33,0.33,0.33,4.52,0.61,0.6,0.58,2.38,0.66,0.62,0.61,2.18,0.26,0.26,0.26,0.345,0.084,0.08,0.081,92.6,3,3.2,3.3,71.5,19,17.8,17.8
\end{filecontents*}
\csvreader[column count = 30, head to column names]{means_and_errorbars_112517.csv}{} 
{$P \ \mathrm{[days]}$ & \POneMean$\pm$\SPSB{\POneUpper}{\POneLower}  &   \PTwoMean$\pm$\SPSB{\PTwoUpper}{\PTwoLower} & \PThreeMean$\pm$\SPSB{\PThreeUpper}{\PThreeLower} & \PFourMean$\pm$\SPSB{\PFourUpper}{\PFourLower} \\
$a \ \mathrm{[AU]}$  & \aOneMean$\pm$\SPSB{\aOneUpper}{\aOneLower}  &   \aTwoMean$\pm$\SPSB{\aTwoUpper}{\aTwoLower} & \aThreeMean$\pm$\SPSB{\aThreeUpper}{\aThreeLower} & \aFourMean$\pm$\SPSB{\aFourUpper}{\aFourLower} \\ 
$K \ \mathrm{[m \ s^{-1}]}$ & \KOneMean$\pm$\SPSB{\KOneUpper}{\KOneLower}  &   \KTwoMean$\pm$\SPSB{\KTwoUpper}{\KTwoLower} & \KThreeMean$\pm$\SPSB{\KThreeUpper}{\KThreeLower} & \KFourMean$\pm$\SPSB{\KFourUpper}{\KFourLower} \\ 
$m \ [M_{\oplus}]$  & \mOneMean$\pm$\SPSB{\mOneUpper}{\mOneLower}  &   \mTwoMean$\pm$\SPSB{\mTwoUpper}{\mTwoLower} & \mThreeMean$\pm$\SPSB{\mThreeUpper}{\mThreeLower} & \mFourMean$\pm$\SPSB{\mFourUpper}{\mFourLower} \\
$e$ & \eOneMean$\pm$\SPSB{\eOneUpper}{\eOneLower}  &  \eTwoMean$\pm$\SPSB{\eTwoUpper}{\eTwoLower} & \eThreeMean$\pm$\SPSB{\eThreeUpper}{\eThreeLower} & \eFourMean$\pm$\SPSB{\eFourUpper}{\eFourLower} \\
$e\cos\omega$ & \ecosomegaOneMean$\pm$\SPSB{\ecosomegaOneUpper}{\ecosomegaOneLower}  &  \ecosomegaTwoMean$\pm$\SPSB{\ecosomegaTwoUpper}{\ecosomegaTwoLower} & \ecosomegaThreeMean$\pm$\SPSB{\ecosomegaThreeUpper}{\ecosomegaThreeLower} & \ecosomegaFourMean$\pm$\SPSB{\ecosomegaFourUpper}{\ecosomegaFourLower} \\
$e\sin\omega$ & \esinomegaOneMean$\pm$\SPSB{\esinomegaOneUpper}{\esinomegaOneLower}  &  \esinomegaTwoMean$\pm$\SPSB{\esinomegaTwoUpper}{\esinomegaTwoLower} & \esinomegaThreeMean$\pm$\SPSB{\esinomegaThreeUpper}{\esinomegaThreeLower} & \esinomegaFourMean$\pm$\SPSB{\esinomegaFourUpper}{\esinomegaFourLower} \\
$i \ [^{\circ}]$ & \multicolumn{4}{c}{\iOneMean$\pm$\SPSB{\iOneUpper}{\iOneLower}} \\
$\Omega \ [^{\circ}]$ & \multicolumn{4}{c}{0.0 (fixed)} \\
$\omega \ [^{\circ}]$ & \argperOneMean$\pm$\SPSB{\argperOneUpper}{\argperOneLower} &   \argperTwoMean$\pm$\SPSB{\argperTwoUpper}{\argperTwoLower} & \argperThreeMean$\pm$\SPSB{\argperThreeUpper}{\argperThreeLower} & \argperFourMean$\pm$\SPSB{\argperFourUpper}{\argperFourLower} \\
$M \ [^{\circ}]$ & \MOneMean$\pm$\SPSB{\MOneUpper}{\MOneLower} &   \MTwoMean$\pm$\SPSB{\MTwoUpper}{\MTwoLower} & \MThreeMean$\pm$\SPSB{\MThreeUpper}{\MThreeLower} & \MFourMean$\pm$\SPSB{\MFourUpper}{\MFourLower} \\ 
$\omega + M \ [^{\circ}]$ & \ArgperPlusMOneMean$\pm$\SPSB{\ArgperPlusMOneUpper}{\ArgperPlusMOneLower}  & \ArgperPlusMTwoMean$\pm$\SPSB{\ArgperPlusMTwoUpper}{\ArgperPlusMTwoLower} & \ArgperPlusMThreeMean$\pm$\SPSB{\ArgperPlusMThreeUpper}{\ArgperPlusMThreeLower} & \ArgperPlusMFourMean$\pm$\SPSB{\ArgperPlusMFourUpper}{\ArgperPlusMFourLower} \\
\hline
\multicolumn{5}{c}{GP covariance parameters} \\ 
\hline
Parameter & Estimate & & \multicolumn{1}{c|}{Parameter} & Estimate \\
\hline
$h  \ \mathrm{[m \ s^{-1}]}$ & \hMean$\pm$\SPSB{\hUpper}{\hLower} & & 
\multicolumn{1}{c|}{$\theta \ \mathrm{[days]}$} & \logProtMean$\pm$\SPSB{\logProtUpper}{\logProtLower}\\
$\lambda \ \mathrm{[days]}$ & \LambdaMean$\pm$\SPSB{\LambdaUpper}{\LambdaLower} & & 
\multicolumn{1}{c|}{$w$} & \wMean$\pm$\SPSB{\wUpper}{\wLower} \\ 
\hline
\multicolumn{5}{c}{Instrumental jitters and offsets (all units of $\mathrm{m \ s^{-1}}$)} \\
\hline
$\sigma_{\mathrm{jit, HIRES_{pre}}}$ & \SigmaJitThreeMean$\pm$\SPSB{\SigmaJitThreeUpper}{\SigmaJitThreeLower} &  & 
\multicolumn{1}{c|}{$\gamma_{\mathrm{HIRES_{pre}}}$} & \offsetThreeMean$\pm$\SPSB{\offsetThreeUpper}{\offsetThreeLower}\\ 
$\sigma_{\mathrm{jit, HIRES_{post}}}$ & \SigmaJitFourMean$\pm$\SPSB{\SigmaJitFourUpper}{\SigmaJitFourLower} &  & 
\multicolumn{1}{c|}{$\gamma_{\mathrm{HIRES_{post}}}$} & \offsetFourMean$\pm$\SPSB{\offsetFourUpper}{\offsetFourLower}\\
$\sigma_{\mathrm{jit, HARPS_{C10}}}$ & \SigmaJitOneMean$\pm$\SPSB{\SigmaJitOneUpper}{\SigmaJitOneLower} &  & 
\multicolumn{1}{c|}{$\gamma_{\mathrm{HARPS_{C10}}}$} & \offsetOneMean$\pm$\SPSB{\offsetOneUpper}{\offsetOneLower}\\ 
$\sigma_{\mathrm{jit, HARPS_{new}}}$ & \SigmaJitTwoMean$\pm$\SPSB{\SigmaJitTwoUpper}{\SigmaJitTwoLower} &  & 
\multicolumn{1}{c|}{$\gamma_{\mathrm{HARPS_{new}}}$} & \offsetTwoMean$\pm$\SPSB{\offsetTwoUpper}{\offsetTwoLower}\\ 
$\sigma_{\mathrm{jit, APF}}$ & \SigmaJitFiveMean$\pm$\SPSB{\SigmaJitFiveUpper}{\SigmaJitFiveLower} &  & 
\multicolumn{1}{c|}{$\gamma_{\mathrm{APF}}$} & \offsetFiveMean$\pm$\SPSB{\offsetFiveUpper}{\offsetFiveLower}\\
$\sigma_{\mathrm{jit, PFS}}$ & \SigmaJitSixMean$\pm$\SPSB{\SigmaJitSixUpper}{\SigmaJitSixLower} &  &  
\multicolumn{1}{c|}{$\gamma_{\mathrm{PFS}}$} & \offsetSixMean$\pm$\SPSB{\offsetSixUpper}{\offsetSixLower}} 
\end{longtable}
\caption{The best-fit planet parameters for the coplanar four-planet fit. The stellar mass is $M_{\star} = 0.37 \ M_{\odot}$ \citep{2016MNRAS.455.2484N}. The osculating Keplerian orbital elements are reported in Jacobi coordinates at the epoch of the first observation, 2450602.09311 BJD. The estimates reported are the posterior distribution means, and the lower and upper uncertainties are 16\textsuperscript{th} and 84\textsuperscript{th} percentiles. Note that before calculating the means and percentiles for $\omega$, $M$, and $\omega + M$, we wrapped their posterior distribution histograms so that the mode of each distribution was at the center. Subsequently, the means were modded to the range $[0^{\circ}, 360^{\circ}]$. In addition, before estimating $\lambda$ and $w$, we eliminated samples from chains that had become stuck in a disfavored local minimum with $\lambda < 20$ days and/or with $w > 0.7$. See the text for more details.  The log-likelihood corresponding to this set of best-fit orbital parameters is $\ln \mathcal{L} = -1404.4$, which will be useful for later comparison.} 
\end{center}
\end{table*}

\subsection{Comparison with a three-planet model}
\cite{2016MNRAS.455.2484N} used Bayesian model selection to compare the RV fits to three, four, and five planet models. They found the strongest evidence for the four-planet model, though the Bayes factor comparison to the model with just the three innermost planets was somewhat borderline. With our updated dataset and Gaussian Process stellar noise model, it is worthwhile to reanalyze the evidence for three or five planet models. The five planet model will be considered in the next section.

Using a model with just the three innermost planets, we find that the maximum log-likelihood of the N-body/GP fit is $\ln \mathcal{L} = -1446$. The difference in the Bayesian Information Criterion (BIC) between the three and four-planet models is $\Delta \mathrm{BIC} \approx 53$, in strong favor of the four-planet model and in agreement with the results of \cite{2016MNRAS.455.2484N}. It is therefore conclusive that the GJ 876 system contains at least four planets.

\section{Fifth Planet Search}

Previous studies have suggested that GJ 876 could potentially host an additional planet in the resonant chain. \cite{2010AA...511A..21C} conducted dynamical stability analyses with a $K = 1$ m/s object with different possible orbital configurations. At that time, only the innermost three planets were known. Their results showed regions of stability beyond 1 AU or at $\sim 0.083$ AU ($P \sim 15$ days), corresponding to a 2:1 interior MMR with planet ``c''. \cite{2012CeMDA.113...35G} numerically integrated the system with all four known planets and also found a region of stability near $P \sim 15$ days. When they integrated one particular five-planet orbital configuration for longer, however, they found that it went unstable before $10^5$ years. 

Before conducting a five planet RV fit with a planet in a $P \sim 15$ day orbit, we must first revisit this question of such a planet's long-term stability and resonant participation.   

\subsection{Fifth planet stability and resonant participation}

Starting with the best-fit parameters from the coplanar, four-planet model (see Section~\ref{four-planet fit}), we conducted a series of N-body integrations with a coplanar, 2.5 $M_{\oplus}$ ($K \approx 1$ m/s) body with a wide range of orbital elements. Each integration was initialized with random orbital elements in the ranges $P_{5^{th}} \in [14.8, 15.2 \ \mathrm{days}]$, $e_{5^{th}} \in [0, 0.3]$, $\omega_{5^{th}} \in [0, 360^{\circ}]$, and $M_{5^{th}} \in [0, 360^{\circ}]$. All integrations were performed with the Wisdom-Holman \texttt{WHFast} code \citep{2015MNRAS.452..376R} that is part of the REBOUND N-body software package \citep{2012A&A...537A.128R}. We used a timestep of 0.15 days, integrated for $10^4$ years, and employed symplectic correctors of order 11.
In order to prevent unnecessary computation, we prematurely terminated any integration in which the $P \sim 15$ day planet underwent orbital instability with $(a - a_0)/a_0 > 0.1$.

\begin{figure}
\epsscale{1.25}
\plotone{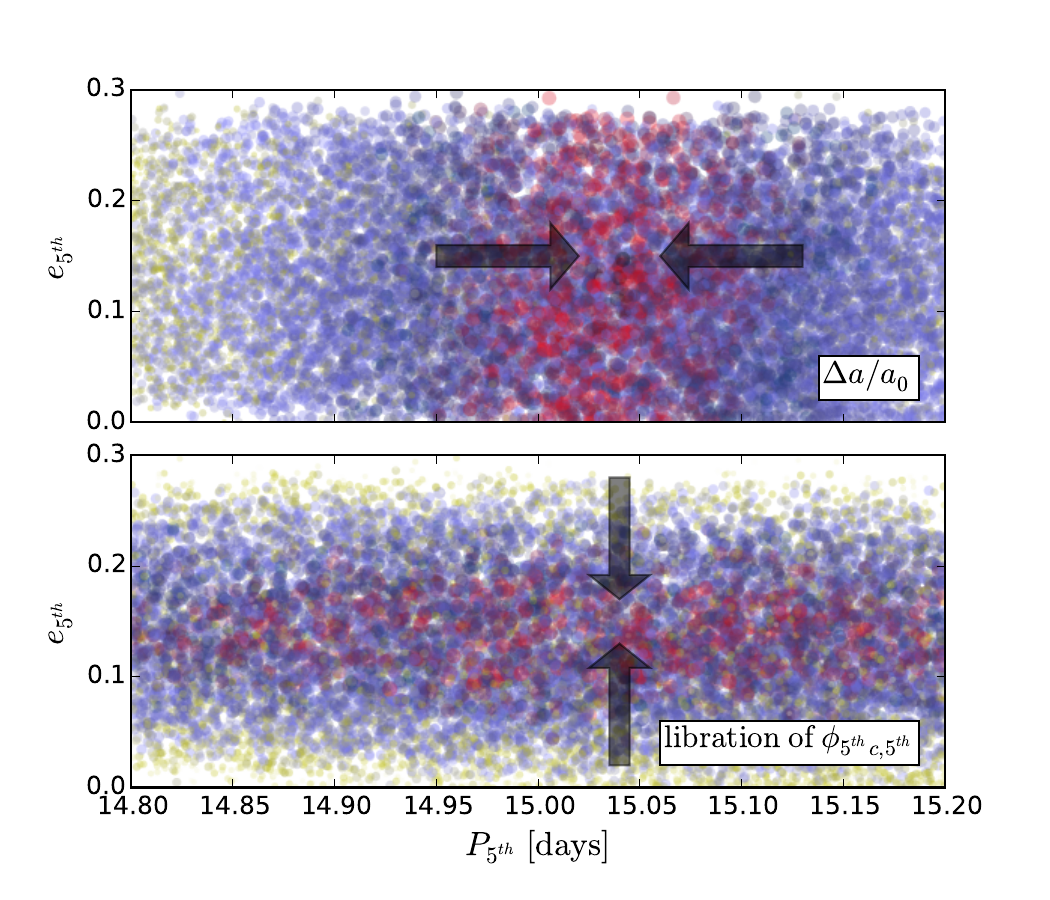}
\caption{The results of the $10^4$ year stability integrations in period/eccentricity space for a potential, 2.5 $M_{\oplus}$ ($K \approx 1$ m/s) planet near $P \sim 15$ days. Only configurations in which the two-body critical resonant angle, $\phi_{5^{th} c, 5^{th}} = 2\lambda_c - \lambda_{5^{th}} - \varpi_{5^{th}}$, was in libration are plotted. In the top panel, the coloration and sizes of the points are according to the test body's maximum fractional semi-major axis deviation from the beginning of the integration. The reddest points (indicated with the arrows) have the smallest $\Delta a/a_0$. In the bottom panel, the point coloration and size reflects the libration amplitude of the two-body critical resonant angle, $\phi_{5^{th} c, 5^{th}} = 2\lambda_c - \lambda_{5^{th}} - \varpi_{5^{th}}$, with the reddest points corresponding to the smallest libration amplitude. In addition to constraints in $P_{5^{th}}$ and $e_{5^{th}}$, there are also strong constraints on the mean longitude. Specifically, in order to be resonance, 
$\lambda_{5^{th}} \in [\sim 140^{\circ},  250^{\circ}] \cup [\sim310^{\circ},  410^{\circ}]$.}
\label{period and eccentricity constraints}
\end{figure}

The test integrations confirm a region of stability and resonant protection for a potential planet in a 2:1 interior MMR with planet ``c''. Figure~\ref{period and eccentricity constraints} shows the integration results projected onto $P_{5^{th}}/e_{5^{th}}$ space. The top panel is colored by semi-major axis stability (specifically, the maximum fractional deviation from the beginning of the integration). The bottom panel is colored by the libration amplitude of the two-body resonant angle, $\phi_{5^{th} c, 5^{th}} = 2\lambda_c - \lambda_{5^{th}} - \varpi_{5^{th}}$. We find that the stable resonant configuration is asymmetric, with $\phi_{5^{th} c, 5^{th}}$ librating around $\pm \sim 40^{\circ}$, and with both $\phi_{5^{th} c, c} = 2\lambda_c - \lambda_{5^{th}} - \varpi_c$ and the three-body angle, $\phi_{\mathrm{Lap}} = \lambda_{5^{th}} - 3\lambda_c + 2\lambda_e$, librating around $\pm \sim 70^{\circ}$. 
The reddest points in the diagram, indicated with the arrows, are the most stable (top panel) or deepest in resonance (bottom panel). Therefore, the orbital configuration that is most favorable has $P_{5^{th}} \sim 15.03$ days and $e_{5^{th}} \sim 0.15$. The test integrations also impose limits on the mean longitude; in order to be in resonance, 
$\lambda_{5^{th}} \in [\sim 140^{\circ},  250^{\circ}]$ or $[\sim310^{\circ},  410^{\circ}]$, where the reference is to the epoch of the fit in Table~\ref{four-planet fit parameters}.

These test simulations -- despite identifying a region of resonant protection/orbital stability -- were only up to $10^4$ years. Is the $P \sim 15$ day body capable of surviving for long periods of time? To test this, we ran a small number additional simulations for $10^7$ years. Most configurations became unstable before the $10^7$ years was up, but many configurations did indeed stay in the stable asymmetric resonant configuration for the duration of the integration. Although $10^7$ years is still a fairly short length of time, these findings suggest that a long-lived, stable resonant configuration is plausible and is worth searching for in the RV data.

\subsection{Five-planet fit}

If the fifth planet is truly there, the RVs should show evidence for it, and the fit should converge in the small region of stable parameter space. We performed a coplanar, five-planet fit to the RV dataset. The Markov Chain initial distributions for the parameters of the four known planets, the covariance parameters, and the instrumental parameters were the same as in Section~\ref{four-planet fit}. The initial distributions of the potential fifth planet's parameters were as follows: $P_{5^{th}} \sim \mathcal{N}(\mu = 15.0 \ \mathrm{days}, \sigma = 0.1, a = 14.8, b = 15.2)$ (a truncated Gaussian), $e_{5^{th}} \sim \mathcal{N}(\mu = 0.15, \sigma = 0.05, a = 0, b = 0.5)$, $\omega_{5^{th}} \sim \mathcal{U}[0, 360^{\circ}]$, and $M_{5^{th}} \sim \mathcal{U}[0, 360^{\circ}]$. Similar to the four-planet fit (Section~\ref{four-planet fit}), we used the Gelman-Rubin diagnostic to monitor chain convergence. 

The results of the five-planet fit do not show evidence for the existence of the additional planet. 
The maximum log-likelihood of the fit was $-1405.6$, compared to $-1404.4$ of the coplanar, four-planet fit (see Table~\ref{four-planet fit parameters}). Even though the five-planet fit has more free parameters, the fact that the maximum $\ln \mathcal{L}_{5\mathrm{pl}.}$ is slightly \textit{smaller} than the maximum $\ln \mathcal{L}_{4\mathrm{pl}.}$ indicates that the five-planet model is not favored. The constraint on $K_{5^{th}}$ from the marginalized posterior distribution was $K_{5^{th}} = 0.54\pm\SPSB{0.45}{0.43}$ m/s. In terms of mass, this corresponds to $m_{5^{th}} = 1.29\pm\SPSB{1.03}{1.00} \ m_{\oplus}$. These estimates represent the approximate maximum RV semi-amplitude/planet mass that we have ruled out with this fit.

The conclusion regarding the potential fifth planet's existence is clear when deduced using the GP regression for stellar activity modeling. Prior to adopting the GP approach, however, our analysis using a white noise model did seem to suggest some evidence for the $P \sim 15$ day planet's existence. 

These contradictory results might be understood as effects arising out of the correlated RV noise from stellar activity. Without the GP model accounting for correlated noise, the extra degrees of freedom of the five-planet model allow some of the noise to be absorbed, making it more favorable than a four-planet model. To the contrary, the extra degrees of freedom are not necessary when the correlated noise is better removed. This interpretation is still speculative. To examine it further, we conducted a series of fits with a set of GJ 876-like synthetic RV data. The results are summarized in the next section. \\

\subsection{Synthetic data fits}
\label{synthetic data fits}


The synthetic data analysis that we will now describe was designed to investigate the behavior of four-planet and five-planet fits to data with only four planets, while considering models that accounted for correlated noise and models that did not. We generated a four-planet, synthetic dataset as follows. We used the best-fit parameters of the four-planet fit (Table~\ref{four-planet fit parameters}) and generated the corresponding N-body model at the timestamps of the real RV dataset. We then perturbed the model data with correlated and uncorrelated noise sources. The synthetic correlated noise was obtained by sampling from the predictive GP model conditioned on the residuals (i.e. the activity-induced RVs) of the four-planet fit. (These residuals were plotted in Figure~\ref{RV residuals}.) The synthetic white noise was generated using the best-fit instrumental jitter parameters. For example, timestamps corresponding to the original APF data were perturbed with Gaussian random variables distributed according to $\mathcal{N}(0, \sigma_{\mathrm{jit, APF}}^2)$. We did not include instrumental zero-point offsets in the synthetic dataset. 

We performed four different MCMC fits to the synthetic dataset, the configurations of which are displayed in Table~\ref{synthetic data fit results}. Here the ``fit type'' row signifies whether or not the Gaussian Process model was used. The ``GP'' fits were conducted as described in Section~\ref{GP}. The ``$\chi_{\mathrm{eff}}^2$'' fits did not use the GP model; instead, the log-likelihood was calculated as $\ln\mathcal{L}(\mathbf{r}|\bm{\theta}) = -\chi_{\mathrm{eff}}^2/2$, with
\begin{equation}
\begin{split}
\chi_{\mathrm{eff}}^2 
&= \sum_{i=1}^{n} \frac{[\mathrm{RV}_{\mathrm{obs}}(t_i) - \mathrm{RV}_{\mathrm{N-body}}(t_i;\bm{\theta})]^2}{\sigma_{\mathrm{obs}}^2(t_i) + \sigma_{\mathrm{jit, inst}}^2} \\
& + \sum_{i=1}^{n}\ln\left[\frac{\sigma_{\mathrm{obs}}^2(t_i) + \sigma_{\mathrm{jit,inst}}^2}{\sigma_{\mathrm{obs}}^2(t_i)}\right].
\end{split}
\end{equation}
This is a commonly-used formulation in which the RV noise is captured with white noise jitter parameters added in quadrature to the reported observational uncertainties. The table reports the maximum log-likelihood of the posterior samples for each of the four different MCMC fits. As before, the Gelman-Rubin diagnostic was used to measure chain convergence.

\setlength{\extrarowheight}{5pt}
\begin{table}[!h]
\caption{Maximum log-likelihoods of the posterior samples from four different fits to the four-planet synthetic data.} 
\label{synthetic data fit results} 
\begin{center}
\begin{tabular}{ c  c  | c  c } 
& & \multicolumn{2}{c}{\# of planets in the fit} \\
& & 4 pl. & 5 pl. \\
\cline{1-4}
\multirow{3}{*}{\rotatebox[origin=c]{90}{   fit type}} & GP & -1409.0 & -1409.6 \\ 
& $\chi_{\mathrm{eff}}^2$ & -385.9 & -376.0 \\
\end{tabular}
\end{center}
\end{table}

For the GP model, $\ln \mathcal{L}_{4\mathrm{pl}.} > \ln \mathcal{L}_{5\mathrm{pl}.}$, as expected. This implies that from the maximum log-likelihoods alone, we can correctly deduce that there is no evidence for five planets in the synthetic data. For the $\chi_{\mathrm{eff}}^2$ model on the other hand, $\ln \mathcal{L}_{4\mathrm{pl}.} < \ln \mathcal{L}_{5\mathrm{pl}.}$. The discrepancy may be explained in that the extra degrees of freedom in the five-planet model allow for a reduction in the synthetic correlated noise, making the fit seem more favorable over the four-planet model. This is the case even though the stellar rotation period is much different from the planet's $P \sim 15$ day period. Only through more advanced techniques -- such as computations of Bayes factors for model selection \citep{2016MNRAS.455.2484N} -- could one determine that the four-planet model is a better fit to the data. 

So far we have only discussed comparisons of the maximum log-likelihoods of the fits. We also investigated some finer details, such as the agreement of the posterior distributions between the GP and $\chi_{\mathrm{eff}}^2$ fits. We compared the marginalized posterior distributions of the 4 planet/GP fit with those of the 4 planet/$\chi_{\mathrm{eff}}^2$ fit. There were no significant differences between the widths of the posterior distributions in the GP fit and those of the $\chi_{\mathrm{eff}}^2$ fit, and the distributions of all parameters were generally consistent within uncertainty. An exception is the set of jitter parameters, which were smaller for the GP fit. This is expected given the structural differences between the models. There were also some interesting differences in the resonant libration amplitude estimates, but that will be discussed in next section.

To summarize the results of our experiments with four-planet synthetic data,  the five-planet fit with a white noise model erroneously yielded a larger log-likelihood compared to the four-planet fit, but this was not the case with the GP correlated noise model. This might explain why the GP model did not show evidence for five planets in the real data, even though preliminary investigations without GPs did. Our results support previous works that show that using a GP model can help prevent spurious planet detections \citep[e.g.][]{2014MNRAS.443.2517H,2015MNRAS.452.2269R}.

It is clear that the four-planet model is the best fit to the data. For the remainder of our analysis, we focus on the details of this model. 

\section{Updated Dynamical Constraints}
\label{updated dynamical constraints}

With our final fit in place, it is of interest to consider how our characterization of the system differs from previous studies. Presumably, our extended RV dataset and GP noise model will have enabled the system parameter estimates to converge in a slightly more accurate and perhaps more stable part of parameter space. \\ \\

\subsection{Libration of the critical resonant angles}
\label{critical resonant angles}

The best probes of the dynamical state of the system are the amplitudes of the various critical resonant angles. Table~\ref{resonant libration amplitudes} displays estimates of the libration amplitudes of the primary resonant angles for the 2:1 MMR between planets ``c'' and ``b'' (first three rows), the 2:1 MMR between ``b'' and ``e'' (second three rows), the 4:1 MMR between ``c'' and ``e'' (next five rows), and the three-body 4:2:1 Laplace resonance involving ``c'', ``b'', and ``e''. 

The libration amplitude estimates in Table~\ref{resonant libration amplitudes} were calculated using the set of converged chains from the coplanar, four-planet fit in Section~\ref{four-planet fit}. We took the parameters of $\sim 7500$ of the posterior samples and integrated forward for $10^5$ years. All integrations were performed with REBOUND's \texttt{WHFast} code \citep{2012A&A...537A.128R, 2015MNRAS.452..376R} with a timestep of 0.15 days and symplectic correctors of order 11. The amplitudes of the critical resonant angles were computed using
\begin{equation}
\mathrm{amp} = \sqrt{\frac{2}{N} \sum_{i=1}^{N} {(\phi_i - \overline{\phi})}^2},
\end{equation}
where $i$ indexes over time, and $\overline{\phi}$ is the libration center. This calculation, which is consistent with that of \cite{2016MNRAS.455.2484N}, is a good approximation to the amplitude of a signal undergoing sinusoidal-type librations. 

\begin{table}[!t]
\caption{Libration centers and amplitudes of the critical resonant angles calculated from $10^5$ year integrations of $\sim 7500$ posterior samples from the coplanar, four-planet fit of Section~\ref{four-planet fit}. The estimates reported are the distribution means, and the lower and upper uncertainties are 16\textsuperscript{th} and 84\textsuperscript{th} percentiles. The fourth column, provided here for convenience, lists the libration amplitudes from the \cite{2016MNRAS.455.2484N} coplanar, four-planet fit (see their Table 8). Entries marked with a ``c'' represent angles that are circulating. } 
\label{resonant libration amplitudes} 
\small
\begin{center}
\begin{tabular}{ c  c  c  c } 
\hline
 \hline
Angle & Cen. & Amp. & Amp.(N16) \\
\hline
$\phi_{cb, c} = 2\lambda_b - \lambda_c - \varpi_c$ & $0^{\circ}$ & 2.5$\pm$\SPSB{0.3}{0.4} & 4.0$\pm$\SPSB{1.6}{1.4}\\
$\phi_{cb, b} = 2\lambda_b - \lambda_c - \varpi_b$  & $0^{\circ}$ & 10.4$\pm$\SPSB{1.7}{1.9} & 13.4$\pm$\SPSB{3.0}{1.4}\\
$\varpi_b - \varpi_c$  & $0^{\circ}$ & 11.9$\pm$\SPSB{1.9}{2.1} & 14.7$\pm$\SPSB{4.0}{3.3}\\ 
\hline
$\phi_{be, b} = 2\lambda_e - \lambda_b - \varpi_b$  & $0^{\circ}$ & 25.0$\pm$\SPSB{4.2}{4.8} & 31.3$\pm$\SPSB{11.9}{8.7}\\
${\phi_{be, e} = 2\lambda_e - \lambda_b - \varpi_e}$\footnote{\label{footnote}These angles sometimes circulate, but they spend much more time in libration.}  & $180^{\circ}$ & 85.9$\pm$\SPSB{8.1}{8.7} & c\\
${\varpi_e - \varpi_b}$\footref{footnote}  & $180^{\circ}$ & 85.0$\pm$\SPSB{8.2}{8.9} & c\\
\hline 
$\phi_{ce, 0} = 4\lambda_e - \lambda_c - 3\varpi_c$  & $0^{\circ}$ & 54.6$\pm$\SPSB{9.2}{10.6} & 67.5$\pm$\SPSB{25.4}{19.4}\\
${\phi_{ce, 1} = 4\lambda_e - \lambda_c - 2\varpi_c - \varpi_e}$\footref{footnote}  & $180^{\circ}$ & 93.5$\pm$\SPSB{9.5}{10.3} & c\\
$\phi_{ce, 2} = 4\lambda_e - \lambda_c - \varpi_c - 2\varpi_e$  &  --  & c & c\\
$\phi_{ce, 3} = 4\lambda_e - \lambda_c - 3\varpi_e$  &  --  & c & c\\
${\varpi_e - \varpi_c}$\footref{footnote} & $180^{\circ}$ &  86.3$\pm$\SPSB{8.5}{9.1} & c\\
\hline
$\phi_\mathrm{Lap} = \lambda_c - 3\lambda_b + 2\lambda_e$ & $0^{\circ}$ & 26.6$\pm$\SPSB{4.4}{5.1} & 33.0$\pm$\SPSB{12.4}{9.3}\\
 \hline
\end{tabular}
\end{center}
\end{table}

Generally speaking, the small libration amplitudes reported in Table~\ref{resonant libration amplitudes} are consistent with the system being quite deep in resonance. The table summarizes the results for all posterior samples as an aggregate population. However, the phase space is actually divided into two separate domains (with a smooth transition between them):
\begin{enumerate}
\item A high energy, quasi-double apsidal corotation resonant (ACR) domain where the angles, $\phi_{be, e}$, $\phi_{ce, 1}$, $\varpi_e - \varpi_b$, and $\varpi_e - \varpi_c$ all primarily librate but undergo brief periods of circulation.
\item A lower energy, stable, double ACR domain where these angles purely librate. The apsidal lines of planets ``c'', ``b'', and ``e'' stay aligned and coprecess at the same average rate.
\end{enumerate}
We will begin by focusing on the first domain -- where the majority ($\sim 90\%$) of posterior samples reside -- and on results applicable to the aggregate population of samples. The second domain will be discussed in detail in the next section. 

Even in the case where $\phi_{be, e}$, $\phi_{ce, 1}$, $\varpi_e - \varpi_b$, and $\varpi_e - \varpi_c$ sometimes circulate, amplitude calculations are still relevant because the angles spend much more time in libration than circulation. They thereby comprise significant terms in the disturbing function expansion. The fact that the secular angles, $\varpi_e - \varpi_b$ and $\varpi_e - \varpi_c$, are nearly always in libration about $180^{\circ}$ indicates that the equilibrium state of the system is a symmetric configuration in which the apsidal line of planet ``e'' quasi-coprecesses with the coprecessing apsidal lines of planets ``c'' and ``b''. In the rare moments in time that $\varpi_e - \varpi_b$ or $\varpi_e - \varpi_c$ circulate, the eccentricity of planet ``e'' reaches a minimum, meaning that the circulation of the periapse is not dynamically significant. The triple conjunctions oscillate about a configuration with planets ``c'' and ``b'' at their periapses and planet ``e'' at its apoapse. 

How does the average dynamical state of the system compare with previous characterizations? Table~\ref{resonant libration amplitudes} also lists the libration amplitude estimates from the four-planet, coplanar fit of \cite{2016MNRAS.455.2484N}, reproduced for convenience from their Table 8. As apparent by comparing our estimates with theirs, the system appears to be deeper in resonance than previously thought; all angles are librating with smaller amplitudes. 

One potential concern in this comparison is that our estimates were calculated using $10^5$ year integrations, while those of \cite{2016MNRAS.455.2484N} used $10^7$ year integrations. If there is systematic drift in the libration amplitudes, that could contribute to the observed differences. To test this, we took 15 of the posterior samples and integrated those systems for $10^7$ years. 
We calculated the resonant amplitudes in 100 $10^5$ year-long segments. The amplitudes drift stochastically but do not appear to do so in a strongly systematic fashion. The mean libration amplitudes resulting from $10^5$ year integrations should be sufficient for comparison.

It does therefore seem to be the case that the system is deeper in resonance than previously thought. This is true in the average sense, when considering posterior samples residing in both of the two domains delineated above. The detailed implications of this will be examined in the next section. First, however, we will take a moment to consider why it is the case that we find the system to be deeper in resonance.

It is possible that the resonant libration amplitudes are lower because we are using a model that accounts for correlated RV noise. We investigated this hypothesis by inspecting the synthetic data fits that we produced in Section~\ref{synthetic data fits}. We took $\sim3500$ posterior samples each from the 4 planet/GP fit and 4 planet/$\chi_{\mathrm{eff}}^2$ fit and measured their resonant libration amplitudes in the same manner as discussed at the beginning of this subsection. Interestingly, we find that the libration amplitude estimates of the $\chi_{\mathrm{eff}}^2$ fit are systematically different than those of the GP fit. The posterior means for the amplitudes of all resonant angles are larger in the $\chi_{\mathrm{eff}}^2$ fit. They are larger by $1\sigma$ on average. Meanwhile, the posterior means of the GP fit are close to the input values of the synthetic data. For the $\chi_{\mathrm{eff}}^2$ fit, there is also a significantly smaller number of posterior samples in the double apsidal corotation domain of phase space compared to the GP fit.

This analysis with synthetic RV data therefore suggests that estimates of the resonant libration amplitudes are sensitive to the choice of the stellar activity model. For our particular choice of model for correlated RV noise, we find smaller resonant libration amplitudes than if we assume the stellar activity noise is uncorrelated. To be fully decisive, however, this study would require tests involving a suite of synthetic datasets with different resonant libration amplitudes.

\subsection{Is GJ 876 at the resonant fixed point?}

As mentioned previously, a subset of our posterior samples are in a lower energy domain of phase space corresponding to a stable double apsidal corotation resonance, in which the angles, $\phi_{be, e}$, $\phi_{ce, 1}$, $\varpi_e - \varpi_b$, and $\varpi_e - \varpi_c$, are purely librating. This configuration of the system is also associated with lower libration amplitudes for all other resonant angles, especially $\phi_{\mathrm{Lap}}$. That is, if we were to select the subset of posterior samples for which $\phi_{\mathrm{Lap}}$ is deepest in resonance, they would all be in the double ACR domain. The existence of these low energy, deeply resonant, double ACR configurations is fully consistent with the phase space mapping of \cite{2016MNRAS.460.1094M}. 

\begin{figure}
\epsscale{1.25}
\plotone{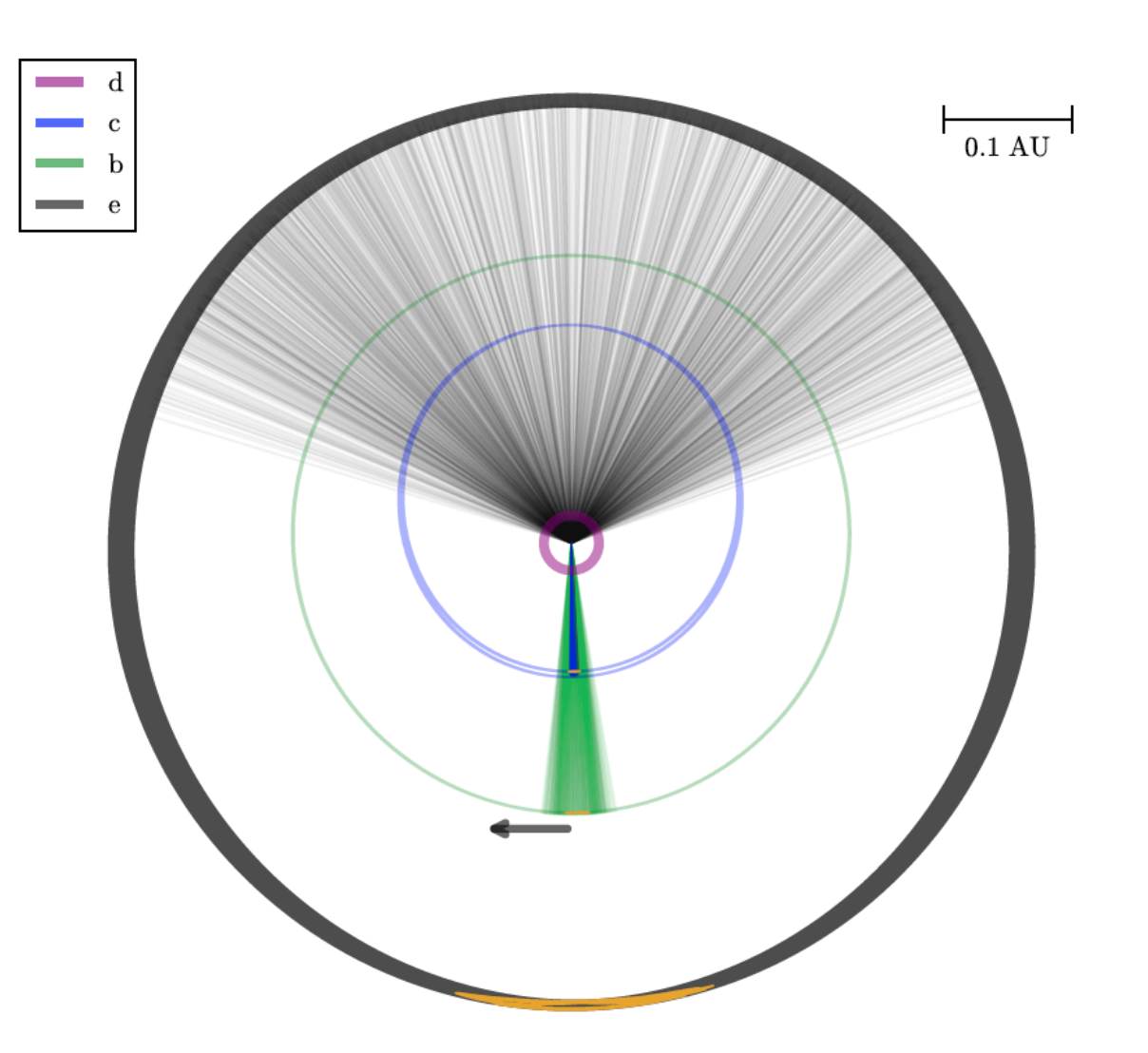}
\caption{An illustration of the GJ 876 orbital evolution in configuration space in a frame that rotates with the shared mean apsidal precession rate ($40.98^{\circ} \mathrm{yr}^{-1}$) of planets ``c'', ``b'', and ``e''. In this depiction, the planets orbit counterclockwise on their orbits, and the orbits themselves regress clockwise, as shown by the left-pointing arrow. The traces of the star-periapse lines are plotted for the outer three planets, illustrating the amplitudes of libration of the apses around perfect corotation. The yellow regions trace the positions of the planets at conjunctions, specifically, the ``c''/``b'' conjunctions for planets ``c'' and ``b'' and the ``b''/``e'' conjunctions for planet ``e''. 
}
\label{configuration space plot}
\end{figure}

Figure~\ref{configuration space plot} displays the orbital evolution of one such posterior sample corresponding to a stable, double ACR configuration. The evolution is plotted in the frame rotating with the mean apsidal precession rate shared by the three resonant planets. The narrowness of the orbits of ``c'' and ``b'', particularly in comparison to those of ``d'' and ``e'', shows how incredibly stable they are. Similarly, the traces of the star-periapse lines and the planet positions at conjunctions (shown in yellow) also convey the depth and stability of the mean-motion and secular apsidal corotation resonances. For this particular orbital evolution, the amplitude of $\phi_\mathrm{Lap}$ is $\sim 10^{\circ}$. We observed with an animation of the orbital evolution in this rotating frame that the precession rates of planets ``b'' and ``e'' tend to be coordinated. For instance, when the apsidal line of planet ``b'' is furthest clockwise, so is that of planet ``e''. The system spends the most time at these extremes.

Which domain of phase space is the system actually in then? The quasi-double ACR or the pure-double ACR (such as that pictured in Figure~\ref{configuration space plot})? Recall from Section~\ref{critical resonant angles} that our updated characterization suggests the average system to be deeper in resonance than determined by \cite{2016MNRAS.455.2484N}. It turns out that this observation is an example of a larger trend that has held true throughout the history of published GJ 876 characterizations. 

\begin{figure}
\epsscale{1.15}
\plotone{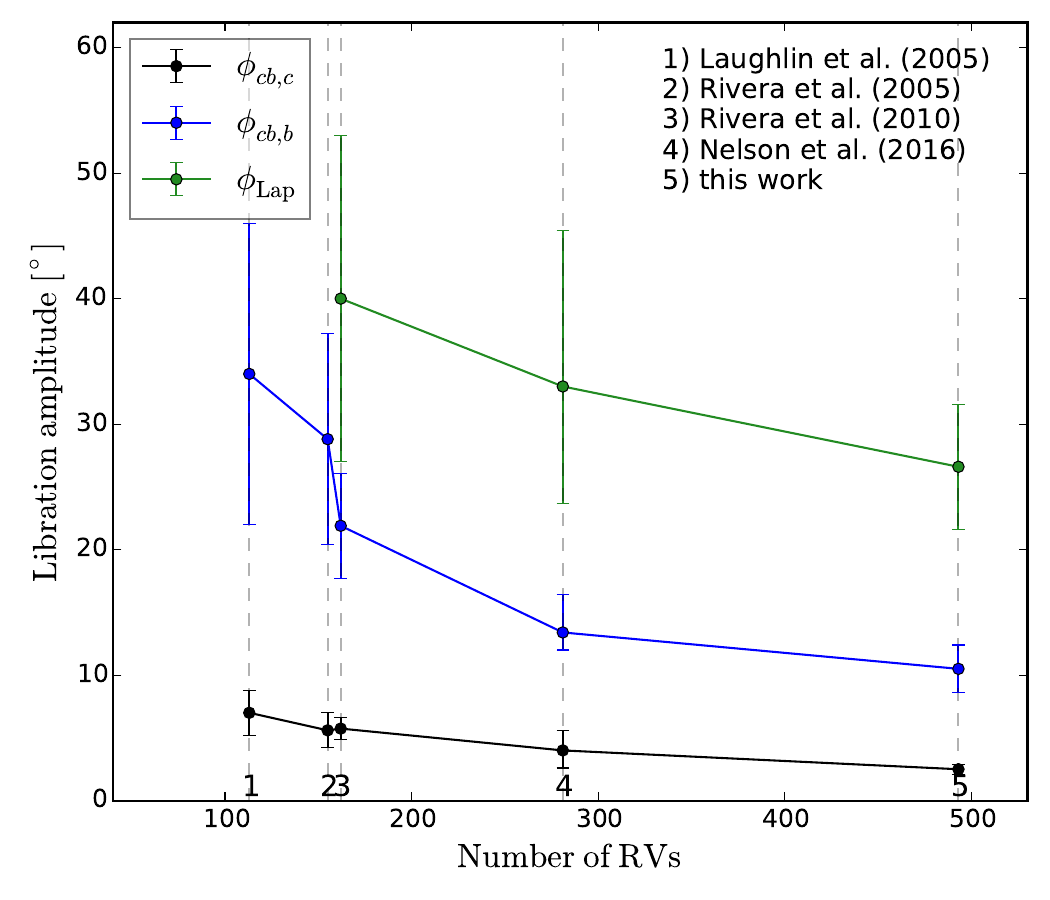}
\caption{The published libration amplitude estimates of three representative critical resonant angles ($\phi_{cb,c}$, $\phi_{cb,b}$, and $\phi_{\mathrm{Lap}}$) as a function of the number of Doppler velocity measurements of the system. These results are taken from \cite{2005ApJ...622.1182L}, \cite{2005ApJ...634..625R}, \cite{2010ApJ...719..890R}, \cite{2016MNRAS.455.2484N}, and this work. The RV count does not include measurements from ELODIE, CORALIE, or Lick Observatory. } 
\label{libration amplitude vs num RVs}
\end{figure}

In Figure~\ref{libration amplitude vs num RVs}, the published literature values of the libration amplitudes of three critical resonant angles are plotted as a function of the number of RV observations contributing to the characterization of the system. The RV count does not include measurements from ELODIE, CORALIE, or Lick Observatory, since observations from Keck, HARPS, and PFS are significantly higher quality. Of course, the number of RV measurements is not the only factor affecting how precise of a system characterization would be available. Data quality, temporal sampling, and baseline are also significant. Here, however, we are simply looking for first-order trends. 

Figure~\ref{libration amplitude vs num RVs} shows that the reported resonant libration amplitudes decreased monotonically as the dataset grew and more precise characterizations became available. In other words, authors have concluded the GJ 876 system to be deeper in resonance upon each successive characterization, prompting the question of whether the true state of the system might be even deeper in resonance than implied by our characterization. However, one must be careful not to read too much into this tentative trend. The studies we are referencing have used a variety of analysis techniques, including both Bayesian and non-Bayesian methods, so the comparison between them is not well-calibrated.

Nevertheless, the apparent trend in literature value libration amplitudes over time is suggestive that the system may be very deep in resonance. It is plausible that GJ 876 is maximally damped and located very close to the resonant fixed point. This possibility is also supported by two findings from the previous section: 1) the existence of deeply-resonant, double ACR configurations among the posterior samples and 2) the potential bias towards larger libration amplitude estimates when using a white noise model as compared to a correlated noise model. Unfortunately, we cannot yet know with certainty which state the system is in. Any effort to converge further would require next-generation RV instruments.

The state of the system and its proximity to the resonant fixed point is also closely linked to the degree of dynamical chaos. This is our next area of exploration. \\ \\

\subsection{Chaotic orbital evolution}

The rapid yet bounded dynamical chaos exhibited by the GJ 876 system is a curiosity that has been examined by several authors \citep{2010ApJ...719..890R, 2013MNRAS.433..928M, 2015AJ....149..167B, 2016MNRAS.455.2484N, 2016MNRAS.460.1094M}. Here we briefly remark on the properties of the chaos suggested by our new characterization of the system. We consider differences between the high energy quasi-double ACR and the low energy pure-double ACR solutions.

\begin{figure}
\epsscale{1.25}
\plotone{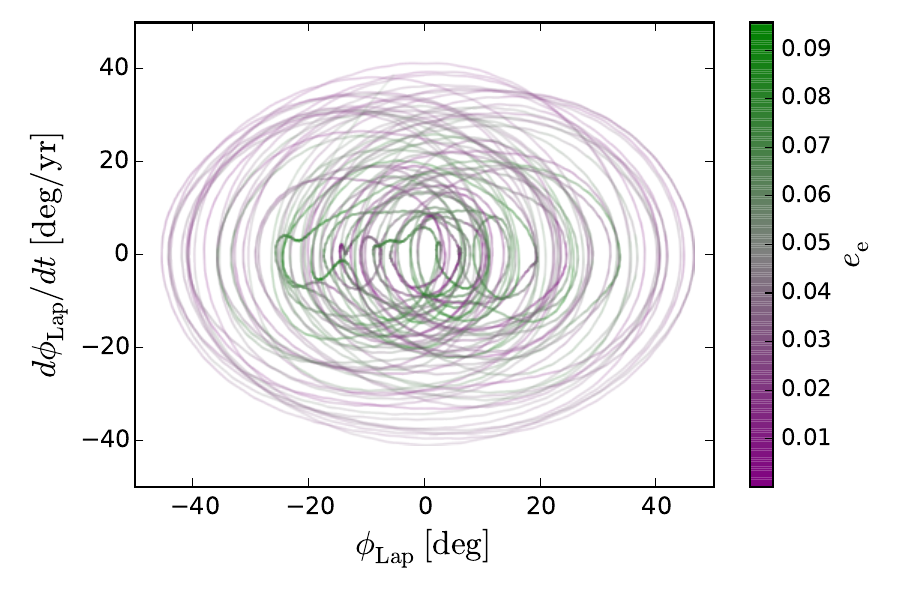}
\caption{Trajectory of a $\sim 500$ year orbital evolution in $d{\phi_{\mathrm{Lap}}}/dt$ vs. $\phi_{\mathrm{Lap}}$ space. The coloration of the line is according to the eccentricity of planet ``e''. The phase space is filled in a short duration of time.} 
\label{Laplace angle trajectory}
\end{figure}

We begin with the high energy configuration. In agreement with previous results, the system's orbital evolution is chaotic on short timescales. This may be illustrated by the ergodicity (filling of phase space) of the trajectory of $\phi_{\mathrm{Lap}}$. In Figure~\ref{Laplace angle trajectory}, we show the time evolution of $d{\phi_{\mathrm{Lap}}}/dt$ vs. $\phi_{\mathrm{Lap}}$ for a $\sim$ 500 year integration of the best-fit four-planet system. The trajectory has been colored by the eccentricity of planet ``e'', and high-frequency oscillations in $\phi_{\mathrm{Lap}}$ have been smoothed using a Savitzky-Golay filter. The trajectory is clearly not periodic and fills phase space in a short duration of time. The maximum excursions in $\phi_{\mathrm{Lap}}$ and $d{\phi_{\mathrm{Lap}}}/dt$ are correlated with small $e_e$, which suggests that the rapid circulation of $\varpi_e$ during minima in $e_e$ is closely linked to the manifestation of the chaos \citep{2015AJ....149..167B, 2016MNRAS.460.1094M}.

Another signature of chaos is repeated crossing of the separatrix in the resonant phase space \citep{1983rsm..book.....L, 2015AJ....149..142F}. We discussed in Section~\ref{critical resonant angles} that the resonant angles $\phi_{be, e}$, $\phi_{ce, 1}$, $\varpi_e - \varpi_b$, and $\varpi_e - \varpi_c$ transition between libration and circulation. This signals that the system does not stay in one domain of phase space, but rather regularly crosses the resonant separatrix. In contrast, the analogous four resonant angles in the Galilean satellite system are all strictly circulating \citep{1986sate.conf..159P}.


Previous characterizations of the GJ 876 system have suggested a startling small Lyapunov timescale on the order of decades \citep{2010ApJ...719..890R, 2011CeMDA.111..235B, 2015AJ....149..167B, 2016MNRAS.455.2484N, 2016MNRAS.460.1094M}. The Lyapunov timescale is a measure of the rate of exponential divergence of nearby trajectories in the phase space of a chaotic system \citep{1999ssd..book.....M}. If two trajectories are separated by a distance $d_0$ at time zero, their separation as a function of time grows as
\begin{equation}
d = d_0 \exp(t/\tau).
\end{equation}
The Lyapunov time, $\tau$, is the time for the displacement to grow by a factor of $e$:
\begin{equation} 
\tau^{-1} = \lim_{t \to \infty}{\frac{\ln(d/d_0)}{t}}.
\end{equation}
The Lyapunov time is related to the maximum Lyapunov characteristic exponent, $\gamma$, by $\tau = \gamma^{-1}$.

Recently, \cite{2016MNRAS.455.2484N} computed a Lyapunov timescale of the system of approximately $\sim 10$ years, consistent with the analytic estimate of \cite{2015AJ....149..167B}. We utilized the maximal Lyapunov characteristic exponent code \citep{2016MNRAS.459.2275R} that is part of the REBOUND N-body software package \citep{2012A&A...537A.128R} to estimate the Lyapunov time for our new characterization of the system. The calculation proceeds by integrating the N-body equations of motion simultaneously with the variational equations, which govern the evolution of small perturbations to an orbit \citep[see e.g.,][]{1983rsm..book.....L, 1999CeMDA..74...59M, 2003PhyD..182..151C, 2010MNRAS.404..837H}. The variational method (as opposed to the two-particle shadow method) is the same approach employed by \cite{2016MNRAS.455.2484N}. 

Considering approximately 300 samples from our best-fit, non-double ACR chains from Section~\ref{four-planet fit}, REBOUND's variational code estimates the Lyapunov time to be $\sim 72\pm\SPSB{28}{13}$ years. This several decades-long estimation is slightly longer than that determined by \cite{2015AJ....149..167B}, \cite{2016MNRAS.455.2484N}, and \cite{2016MNRAS.460.1094M}, but it is not significantly inconsistent. For systems in the low energy double ACR, the Lyapunov time is significantly longer and more variable, $\log_{10}\tau [\mathrm{years}]= 3.7\pm\SPSB{3.4}{1.1}$. This is consistent with the findings of \cite{2015AJ....149..167B} and \cite{2016MNRAS.460.1094M}, who identified the existence of significantly less chaotic configurations in phase space. 

The nature of the dynamical chaos in the system, including its magnitude and characteristic timescale, is closely linked to theories of the system's formation. \cite{2015AJ....149..167B} examined potential assembly scenarios of a chaotic Laplace resonance through stochastic, convergent migration. In the case of a chaotic system with non-negligible libration amplitudes, they showed that the gas disk must be sufficiently turbulent so as to prevent the system from settling into a quasi-periodic, maximally damped configuration. At the same time, however, the disk must be laminar enough so as to not inhibit resonant capture.  Given that both high energy, vigorously chaotic and low energy, more regular solutions are consistent with the RV data, there are significant limits on our ability to say much about the magnitude of turbulence that was present in the primordial disk.

\section{Summary}

Within the current planetary census, few exoplanet systems are better suited to the study of resonance (and specifically, the formation of resonant chains) than Gliese 876. The system exhibits extreme resonant perturbations on readily observable timescales; this owes to the system's large planet-star mass ratio\footnote{The ratio of the sum of the planet masses in GJ 876 to the stellar mass is $\sim 0.009$. By comparison, the ratio for Kepler-223, a system containing a four-planet resonant chain, is $\sim 6.8 \times 10^{-5}$, or more than two orders of magnitude smaller.}, the low-order MMRs, and the significant eccentricities. Moreover, resonant systems like GJ 876 contain clues to the early disk migration processes required to form them, thereby helping illuminate some aspects of this still highly unconstrained mechanism. It is therefore worthwhile to maintain an up-to-date characterization of this exemplary planetary system.

We presented six years of Keck/HIRES, HARPS, APF, and PFS radial velocity measurements of GJ 876, which collectively more than doubled the size of the existing dataset. We employed a complex RV fitting procedure -- complete with a computationally efficient Wisdom-Holman symplectic N-body integrator and a Gaussian Process model for correlated stellar activity-induced RV noise -- and derived new estimates of the system parameters. Throughout our RV modeling, we used the assumption of a coplanar system architecture, which was supported by previous work that determined that the mutual orbital inclinations must be very low to ensure long-term stability \citep{2016MNRAS.455.2484N}. 

The posterior samples are available at \href{https://github.com/smillholland/GJ876}{https://github.com/smillholland/GJ876} and archived at \href{https://doi.org/10.5281/zenodo.1149601}{10.5281/zenodo.1149601}. The samples were not explicitly checked for long-term stability. Performing long-term dynamical integrations may permit further constraint on the system parameters \citep{2016MNRAS.455.2484N,2016ApJ...832L..22T}. Among the new parameter estimates, of particular interest is a significantly smaller value for the eccentricity of the innermost planet; the elevated eccentricity estimates ($\gtrsim 0.1$) of previous characterizations had hitherto been a challenge to explain given the short tidal circularization time expected for such a close-in planet.  

The system appears capable of hosting a fifth planet with $P \sim 15$ days in a long-lived, asymmetric resonant configuration with $\phi_{5^{th} c, c} = 2\lambda_c - \lambda_{5^{th}} - \varpi_c$ and the three-body Laplace angle, $\phi_{\mathrm{Lap}} = \lambda_{5^{th}} - 3\lambda_c + 2\lambda_e$, both librating about $\pm 70^{\circ}$. The RVs, however, suggest no evidence for such a planet. As part of our fifth planet search, we performed experiments with synthetic RV data. The results underscore the importance of properly accounting for stellar activity-induced RVs, even if the magnitude of the noise is not very large.  

We find that the system is decidedly deeper in resonance than previous studies have suggested; all of the critical resonant angles are librating with smaller amplitudes. We surmise that this results from the expanded RV dataset and the more detailed handling of correlated stellar noise. It is possible that this a general rule that applies to other resonant systems as well. This possibility is supported by our experiments with synthetic RV data. We found that a fit to synthetic data that used a white noise model resulted in systematically larger libration amplitude estimates compared to a fit that used a correlated noise model.

Moreover, as the RV dataset has grown over time, the published libration amplitude estimates have decreased monotonically (Figure~\ref{libration amplitude vs num RVs}). This may suggest that the system is even deeper in resonance than implied by our system characterization, since imperfect accounting of stellar and measurement noise could plausibly manifest as inflation of the libration amplitude estimates. The prospects of fully converging, however, are likely only possible with next generation RV instruments and with even more advanced stellar noise handling.   

Our posterior samples are consistent with two different configurations in phase space. The first domain, where $\sim 90\%$ of the posterior samples reside, is a high energy, quasi-double apsidal corotation resonance (ACR), in which the apsidal line of planet ``e'' quasi-coprecesses with the coprecessing apsidal lines of planets ``c'' and ``b''. The secular angles, $\varpi_e - \varpi_c$ and $\varpi_e - \varpi_b$, are nearly always librating about $180^{\circ}$ and circulate when $e_e$ is minimized. The second domain is a low energy, pure double ACR, where the libration amplitudes of all critical angles are small, and the apsidal lines of planets ``c'', ``b'', and ``e'' are corotating. This region is also more stable, has a much longer Lyapunov timescale, and is associated with a significantly smaller libration amplitude for the Laplace angle. 

The RV data does not yet have the precision to reveal with certainty whether the system is in the quasi-regular or more vigorously chaotic domain. However, the observed trend of the literature value libration amplitude estimates over time is suggestive that the system might be nearly maximally damped and close to the fixed point in the resonant phase space. The depth of the system within the resonant potential well is not merely an esoteric intrigue. It has important implications for understanding the environment of the primordial disk, since the amplitude of the resonant librations and the degree of chaos are probes of turbulence during the system's early dissipative evolution. Gliese 876 is by many accounts the archetype of known resonant planetary systems. In this sense, it offers one of the best avenues for illuminating the conditions under which planets can arise. 


\section{Acknowledgements}

We express gratitude to the anonymous referee, whose very thorough review helped improve this paper. We thank Hanno Rein, Dan Tamayo, Fabio Del Sordo, Rapha\"{e}lle Haywood, and Allen Davis for useful discussions. We also acknowledge Guillem Anglada-Escud\'{e} for his assistance with the HARPS-TERRA pipeline. S.M. is supported by the National Science Foundation Graduate Research Fellowship Program under Grant Number DGE-1122492. This material is also based upon work supported by the National Aeronautics and Space Administration through the NASA Astrobiology Institute under Cooperative Agreement Notice NNH13ZDA017C issued through the Science Mission Directorate. We acknowledge support from the NASA Astrobiology Institute through a cooperative agreement between NASA Ames Research Center and Yale University. 

This work made use of the REBOUND code, which is publicly available at \href{http://github.com/hannorein/rebound}{\url{http://github.com/hannorein/rebound}}. We also acknowledge the Yale Center for Research Computing for the use of its high performance computing clusters.

\bibliographystyle{apj}
\bibliography{GJ876_Observations.bbl}

\begin{thebibliography}{}
\expandafter\ifx\csname natexlab\endcsname\relax\def\natexlab#1{#1}\fi

\bibitem[{{Affer} {et~al.}(2016){Affer}, {Micela}, {Damasso}, {Perger},
  {Ribas}, {Su{\'a}rez Mascare{\~n}o}, {Gonz{\'a}lez Hern{\'a}ndez}, {Rebolo},
  {Poretti}, {Maldonado}, {Leto}, {Pagano}, {Scandariato}, {Zanmar Sanchez},
  {Sozzetti}, {Bonomo}, {Malavolta}, {Morales}, {Rosich}, {Bignamini},
  {Gratton}, {Velasco}, {Cenadelli}, {Claudi}, {Cosentino}, {Desidera},
  {Giacobbe}, {Herrero}, {Lafarga}, {Lanza}, {Molinari}, \&
  {Piotto}}]{2016A&A...593A.117A}
{Affer}, L., {Micela}, G., {Damasso}, M., {et~al.} 2016, \aap, 593, A117

\bibitem[{{Ambikasaran} {et~al.}(2015){Ambikasaran}, {Foreman-Mackey},
  {Greengard}, {Hogg}, \& {O'Neil}}]{2015ITPAM..38..252A}
{Ambikasaran}, S., {Foreman-Mackey}, D., {Greengard}, L., {Hogg}, D.~W., \&
  {O'Neil}, M. 2015, IEEE Transactions on Pattern Analysis and Machine
  Intelligence, 38, arXiv:1403.6015

\bibitem[{{Anglada-Escud{\'e}} \& {Butler}(2012)}]{2012ApJS..200...15A}
{Anglada-Escud{\'e}}, G., \& {Butler}, R.~P. 2012, \apjs, 200, 15

\bibitem[{{Angus} {et~al.}(2018){Angus}, {Morton}, {Aigrain}, {Foreman-Mackey},
  \& {Rajpaul}}]{2018MNRAS.474.2094A}
{Angus}, R., {Morton}, T., {Aigrain}, S., {Foreman-Mackey}, D., \& {Rajpaul},
  V. 2018, \mnras, 474, 2094

\bibitem[{{Astudillo-Defru} {et~al.}(2017){Astudillo-Defru}, {D{\'{\i}}az},
  {Bonfils}, {Almenara}, {Delisle}, {Bouchy}, {Delfosse}, {Forveille}, {Lovis},
  {Mayor}, {Murgas}, {Pepe}, {Santos}, {S{\'e}gransan}, {Udry}, \&
  {W{\"u}nsche}}]{2017A&A...605L..11A}
{Astudillo-Defru}, N., {D{\'{\i}}az}, R.~F., {Bonfils}, X., {et~al.} 2017,
  \aap, 605, L11

\bibitem[{{Baluev}(2011)}]{2011CeMDA.111..235B}
{Baluev}, R.~V. 2011, Celestial Mechanics and Dynamical Astronomy, 111, 235

\bibitem[{{Batygin} {et~al.}(2015){Batygin}, {Deck}, \&
  {Holman}}]{2015AJ....149..167B}
{Batygin}, K., {Deck}, K.~M., \& {Holman}, M.~J. 2015, \aj, 149, 167

\bibitem[{{Bean} \& {Seifahrt}(2009)}]{2009AA...496..249B}
{Bean}, J.~L., \& {Seifahrt}, A. 2009, \aap, 496, 249

\bibitem[{{Beaug{\'e}} {et~al.}(2003){Beaug{\'e}}, {Ferraz-Mello}, \&
  {Michtchenko}}]{2003ApJ...593.1124B}
{Beaug{\'e}}, C., {Ferraz-Mello}, S., \& {Michtchenko}, T.~A. 2003, \apj, 593,
  1124

\bibitem[{{Beaug{\'e}} \& {Michtchenko}(2003)}]{2003MNRAS.341..760B}
{Beaug{\'e}}, C., \& {Michtchenko}, T.~A. 2003, \mnras, 341, 760

\bibitem[{{Benedict} {et~al.}(2002){Benedict}, {McArthur}, {Forveille},
  {Delfosse}, {Nelan}, {Butler}, {Spiesman}, {Marcy}, {Goldman}, {Perrier},
  {Jefferys}, \& {Mayor}}]{2002ApJ...581L.115B}
{Benedict}, G.~F., {McArthur}, B.~E., {Forveille}, T., {et~al.} 2002, \apjl,
  581, L115

\bibitem[{{Butler} {et~al.}(1996){Butler}, {Marcy}, {Williams}, {McCarthy},
  {Dosanjh}, \& {Vogt}}]{1996PASP..108..500B}
{Butler}, R.~P., {Marcy}, G.~W., {Williams}, E., {et~al.} 1996, \pasp, 108, 500

\bibitem[{{Cincotta} {et~al.}(2003){Cincotta}, {Giordano}, \&
  {Sim{\'o}}}]{2003PhyD..182..151C}
{Cincotta}, P.~M., {Giordano}, C.~M., \& {Sim{\'o}}, C. 2003, Physica D
  Nonlinear Phenomena, 182, 151

\bibitem[{{Correia} {et~al.}(2010){Correia}, {Couetdic}, {Laskar}, {Bonfils},
  {Mayor}, {Bertaux}, {Bouchy}, {Delfosse}, {Forveille}, {Lovis}, {Pepe},
  {Perrier}, {Queloz}, \& {Udry}}]{2010AA...511A..21C}
{Correia}, A.~C.~M., {Couetdic}, J., {Laskar}, J., {et~al.} 2010, \aap, 511,
  A21

\bibitem[{{Crida} {et~al.}(2008){Crida}, {S{\'a}ndor}, \&
  {Kley}}]{2008A&A...483..325C}
{Crida}, A., {S{\'a}ndor}, Z., \& {Kley}, W. 2008, \aap, 483, 325

\bibitem[{{Damasso} \& {Del Sordo}(2017)}]{2017AA...599A.126D}
{Damasso}, M., \& {Del Sordo}, F. 2017, \aap, 599, A126

\bibitem[{{Delfosse} {et~al.}(1998){Delfosse}, {Forveille}, {Mayor}, {Perrier},
  {Naef}, \& {Queloz}}]{1998A&A...338L..67D}
{Delfosse}, X., {Forveille}, T., {Mayor}, M., {et~al.} 1998, \aap, 338, L67

\bibitem[{{Faria} {et~al.}(2016){Faria}, {Haywood}, {Brewer}, {Figueira},
  {Oshagh}, {Santerne}, \& {Santos}}]{2016A&A...588A..31F}
{Faria}, J.~P., {Haywood}, R.~D., {Brewer}, B.~J., {et~al.} 2016, \aap, 588,
  A31

\bibitem[{{Ford}(2005)}]{2005AJ....129.1706F}
{Ford}, E.~B. 2005, \aj, 129, 1706

\bibitem[{{Foreman-Mackey} {et~al.}(2013){Foreman-Mackey}, {Hogg}, {Lang}, \&
  {Goodman}}]{2013PASP..125..306F}
{Foreman-Mackey}, D., {Hogg}, D.~W., {Lang}, D., \& {Goodman}, J. 2013, \pasp,
  125, 306

\bibitem[{{French} {et~al.}(2015){French}, {Dawson}, \&
  {Showalter}}]{2015AJ....149..142F}
{French}, R.~G., {Dawson}, R.~I., \& {Showalter}, M.~R. 2015, \aj, 149, 142

\bibitem[{Gelman {et~al.}(2014)Gelman, Carlin, Stern, Dunson, Vehtari, \&
  Rubin}]{BDA}
Gelman, A., Carlin, J., Stern, H., {et~al.} 2014, Bayesian Data Analysis, Third
  Edition (Chapman \& {Hall/CRC} Texts in Statistical Science), 3rd edn.
  (London: Chapman and Hall/CRC)

\bibitem[{{Gelman} \& {Rubin}(1992)}]{1992StaSc...7..457G}
{Gelman}, A., \& {Rubin}, D.~B. 1992, Statistical Science, 7, 457

\bibitem[{{Gerlach} \& {Haghighipour}(2012)}]{2012CeMDA.113...35G}
{Gerlach}, E., \& {Haghighipour}, N. 2012, Celestial Mechanics and Dynamical
  Astronomy, 113, 35

\bibitem[{{Go{\'z}dziewski} {et~al.}(2002){Go{\'z}dziewski}, {Bois}, \&
  {Maciejewski}}]{2002MNRAS.332..839G}
{Go{\'z}dziewski}, K., {Bois}, E., \& {Maciejewski}, A.~J. 2002, \mnras, 332,
  839

\bibitem[{{Grunblatt} {et~al.}(2015){Grunblatt}, {Howard}, \&
  {Haywood}}]{2015ApJ...808..127G}
{Grunblatt}, S.~K., {Howard}, A.~W., \& {Haywood}, R.~D. 2015, \apj, 808, 127

\bibitem[{{Haghighipour} {et~al.}(2003){Haghighipour}, {Couetdic}, {Varadi}, \&
  {Moore}}]{2003ApJ...596.1332H}
{Haghighipour}, N., {Couetdic}, J., {Varadi}, F., \& {Moore}, W.~B. 2003, \apj,
  596, 1332

\bibitem[{{Haywood} {et~al.}(2014){Haywood}, {Collier Cameron}, {Queloz},
  {Barros}, {Deleuil}, {Fares}, {Gillon}, {Lanza}, {Lovis}, {Moutou}, {Pepe},
  {Pollacco}, {Santerne}, {S{\'e}gransan}, \& {Unruh}}]{2014MNRAS.443.2517H}
{Haywood}, R.~D., {Collier Cameron}, A., {Queloz}, D., {et~al.} 2014, \mnras,
  443, 2517

\bibitem[{{Hinse} {et~al.}(2010){Hinse}, {Christou}, {Alvarellos}, \&
  {Go{\'z}dziewski}}]{2010MNRAS.404..837H}
{Hinse}, T.~C., {Christou}, A.~A., {Alvarellos}, J.~L.~A., \&
  {Go{\'z}dziewski}, K. 2010, \mnras, 404, 837

\bibitem[{{Ji} {et~al.}(2002){Ji}, {Li}, \& {Liu}}]{2002ApJ...572.1041J}
{Ji}, J., {Li}, G., \& {Liu}, L. 2002, \apj, 572, 1041

\bibitem[{{Jones} {et~al.}(2001){Jones}, {Sleep}, \&
  {Chambers}}]{2001A&A...366..254J}
{Jones}, B.~W., {Sleep}, P.~N., \& {Chambers}, J.~E. 2001, \aap, 366, 254

\bibitem[{{Kammer} {et~al.}(2014){Kammer}, {Knutson}, {Howard}, {Laughlin},
  {Deming}, {Todorov}, {Desert}, {Agol}, {Burrows}, {Fortney}, {Showman}, \&
  {Lewis}}]{2014ApJ...781..103K}
{Kammer}, J.~A., {Knutson}, H.~A., {Howard}, A.~W., {et~al.} 2014, \apj, 781,
  103

\bibitem[{{Kinoshita} \& {Nakai}(2001)}]{2001PASJ...53L..25K}
{Kinoshita}, H., \& {Nakai}, H. 2001, \pasj, 53, L25

\bibitem[{{Kley} {et~al.}(2005){Kley}, {Lee}, {Murray}, \&
  {Peale}}]{2005A&A...437..727K}
{Kley}, W., {Lee}, M.~H., {Murray}, N., \& {Peale}, S.~J. 2005, \aap, 437, 727

\bibitem[{{Kley} {et~al.}(2004){Kley}, {Peitz}, \&
  {Bryden}}]{2004A&A...414..735K}
{Kley}, W., {Peitz}, J., \& {Bryden}, G. 2004, \aap, 414, 735

\bibitem[{{Laughlin} {et~al.}(2005){Laughlin}, {Butler}, {Fischer}, {Marcy},
  {Vogt}, \& {Wolf}}]{2005ApJ...622.1182L}
{Laughlin}, G., {Butler}, R.~P., {Fischer}, D.~A., {et~al.} 2005, \apj, 622,
  1182

\bibitem[{{Laughlin} \& {Chambers}(2001)}]{2001ApJ...551L.109L}
{Laughlin}, G., \& {Chambers}, J.~E. 2001, \apjl, 551, L109

\bibitem[{{Lee} \& {Peale}(2002)}]{2002ApJ...567..596L}
{Lee}, M.~H., \& {Peale}, S.~J. 2002, \apj, 567, 596

\bibitem[{{Lee} \& {Thommes}(2009)}]{2009ApJ...702.1662L}
{Lee}, M.~H., \& {Thommes}, E.~W. 2009, \apj, 702, 1662

\bibitem[{Lichtenberg \& Lieberman(1989)}]{1983rsm..book.....L}
Lichtenberg, A., \& Lieberman, M. 1989, Regular and Stochastic Motion, Applied
  Mathematical Sciences (Springer New York)

\bibitem[{{L{\'o}pez-Morales} {et~al.}(2016){L{\'o}pez-Morales}, {Haywood},
  {Coughlin}, {Zeng}, {Buchhave}, {Giles}, {Affer}, {Bonomo}, {Charbonneau},
  {Collier Cameron}, {Consentino}, {Dressing}, {Dumusque}, {Figueira},
  {Fiorenzano}, {Harutyunyan}, {Johnson}, {Latham}, {Lopez}, {Lovis},
  {Malavolta}, {Mayor}, {Micela}, {Molinari}, {Mortier}, {Motalebi},
  {Nascimbeni}, {Pepe}, {Phillips}, {Piotto}, {Pollacco}, {Queloz}, {Rice},
  {Sasselov}, {Segransan}, {Sozzetti}, {Udry}, {Vanderburg}, \&
  {Watson}}]{2016AJ....152..204L}
{L{\'o}pez-Morales}, M., {Haywood}, R.~D., {Coughlin}, J.~L., {et~al.} 2016,
  \aj, 152, 204

\bibitem[{{Marcy} {et~al.}(2001){Marcy}, {Butler}, {Fischer}, {Vogt},
  {Lissauer}, \& {Rivera}}]{2001ApJ...556..296M}
{Marcy}, G.~W., {Butler}, R.~P., {Fischer}, D., {et~al.} 2001, \apj, 556, 296

\bibitem[{{Marcy} {et~al.}(1998){Marcy}, {Butler}, {Vogt}, {Fischer}, \&
  {Lissauer}}]{1998ApJ...505L.147M}
{Marcy}, G.~W., {Butler}, R.~P., {Vogt}, S.~S., {Fischer}, D., \& {Lissauer},
  J.~J. 1998, \apjl, 505, L147

\bibitem[{{Mart{\'{\i}}} {et~al.}(2016){Mart{\'{\i}}}, {Cincotta}, \&
  {Beaug{\'e}}}]{2016MNRAS.460.1094M}
{Mart{\'{\i}}}, J.~G., {Cincotta}, P.~M., \& {Beaug{\'e}}, C. 2016, \mnras,
  460, 1094

\bibitem[{{Mart{\'{\i}}} {et~al.}(2013){Mart{\'{\i}}}, {Giuppone}, \&
  {Beaug{\'e}}}]{2013MNRAS.433..928M}
{Mart{\'{\i}}}, J.~G., {Giuppone}, C.~A., \& {Beaug{\'e}}, C. 2013, \mnras,
  433, 928

\bibitem[{{Mikkola} \& {Innanen}(1999)}]{1999CeMDA..74...59M}
{Mikkola}, S., \& {Innanen}, K. 1999, Celestial Mechanics and Dynamical
  Astronomy, 74, 59

\bibitem[{{Mortier} {et~al.}(2016){Mortier}, {Faria}, {Santos}, {Rajpaul},
  {Figueira}, {Boisse}, {Collier Cameron}, {Dumusque}, {Lo Curto}, {Lovis},
  {Mayor}, {Melo}, {Pepe}, {Queloz}, {Santerne}, {S{\'e}gransan}, {Sousa},
  {Sozzetti}, \& {Udry}}]{2016A&A...585A.135M}
{Mortier}, A., {Faria}, J.~P., {Santos}, N.~C., {et~al.} 2016, \aap, 585, A135

\bibitem[{Murray \& Dermott(1998)}]{1999ssd..book.....M}
Murray, C.~D., \& Dermott, S.~F. 1998, Solar System Dynamics (Cambridge
  University Press)

\bibitem[{{Murray} {et~al.}(2002){Murray}, {Paskowitz}, \&
  {Holman}}]{2002ApJ...565..608M}
{Murray}, N., {Paskowitz}, M., \& {Holman}, M. 2002, \apj, 565, 608

\bibitem[{{Nelson} {et~al.}(2014){Nelson}, {Ford}, \&
  {Payne}}]{2014ApJS..210...11N}
{Nelson}, B., {Ford}, E.~B., \& {Payne}, M.~J. 2014, \apjs, 210, 11

\bibitem[{{Nelson} {et~al.}(2016){Nelson}, {Robertson}, {Payne}, {Pritchard},
  {Deck}, {Ford}, {Wright}, \& {Isaacson}}]{2016MNRAS.455.2484N}
{Nelson}, B.~E., {Robertson}, P.~M., {Payne}, M.~J., {et~al.} 2016, \mnras,
  455, 2484

\bibitem[{{Peale}(1976)}]{1976ARA&A..14..215P}
{Peale}, S.~J. 1976, \araa, 14, 215

\bibitem[{{Peale}(1986)}]{1986sate.conf..159P}
{Peale}, S.~J. 1986, in Satellites, ed. J.~A. {Burns} \& M.~S. {Matthews},
  159--223

\bibitem[{Press {et~al.}(1992)Press, Teukolsky, Vetterling, \&
  Flannery}]{1992nrfa.book.....P}
Press, W.~H., Teukolsky, S.~A., Vetterling, W.~T., \& Flannery, B.~P. 1992,
  Numerical Recipes in FORTRAN (2nd Ed.): The Art of Scientific Computing (New
  York, NY, USA: Cambridge University Press)

\bibitem[{{Rajpaul} {et~al.}(2015){Rajpaul}, {Aigrain}, {Osborne}, {Reece}, \&
  {Roberts}}]{2015MNRAS.452.2269R}
{Rajpaul}, V., {Aigrain}, S., {Osborne}, M.~A., {Reece}, S., \& {Roberts}, S.
  2015, \mnras, 452, 2269

\bibitem[{Rasmussen \& Williams(2005)}]{2006gpml.book.....R}
Rasmussen, C.~E., \& Williams, C. K.~I. 2005, Gaussian Processes for Machine
  Learning (Adaptive Computation and Machine Learning) (The MIT Press)

\bibitem[{{Rein} \& {Liu}(2012)}]{2012A&A...537A.128R}
{Rein}, H., \& {Liu}, S.-F. 2012, \aap, 537, A128

\bibitem[{{Rein} \& {Tamayo}(2015)}]{2015MNRAS.452..376R}
{Rein}, H., \& {Tamayo}, D. 2015, \mnras, 452, 376

\bibitem[{{Rein} \& {Tamayo}(2016)}]{2016MNRAS.459.2275R}
---. 2016, \mnras, 459, 2275

\bibitem[{{Rivera} {et~al.}(2010){Rivera}, {Laughlin}, {Butler}, {Vogt},
  {Haghighipour}, \& {Meschiari}}]{2010ApJ...719..890R}
{Rivera}, E.~J., {Laughlin}, G., {Butler}, R.~P., {et~al.} 2010, \apj, 719, 890

\bibitem[{{Rivera} \& {Lissauer}(2001)}]{2001ApJ...558..392R}
{Rivera}, E.~J., \& {Lissauer}, J.~J. 2001, \apj, 558, 392

\bibitem[{{Rivera} {et~al.}(2005){Rivera}, {Lissauer}, {Butler}, {Marcy},
  {Vogt}, {Fischer}, {Brown}, {Laughlin}, \& {Henry}}]{2005ApJ...634..625R}
{Rivera}, E.~J., {Lissauer}, J.~J., {Butler}, R.~P., {et~al.} 2005, \apj, 634,
  625

\bibitem[{{Shankland} {et~al.}(2006){Shankland}, {Rivera}, {Laughlin}, {Blank},
  {Price}, {Gary}, {Bissinger}, {Ringwald}, {White}, {Henry}, {McGee}, {Wolf},
  {Carter}, {Lee}, {Biggs}, {Monard}, \& {Ashley}}]{2006ApJ...653..700S}
{Shankland}, P.~D., {Rivera}, E.~J., {Laughlin}, G., {et~al.} 2006, \apj, 653,
  700

\bibitem[{{Snellgrove} {et~al.}(2001){Snellgrove}, {Papaloizou}, \&
  {Nelson}}]{2001A&A...374.1092S}
{Snellgrove}, M.~D., {Papaloizou}, J.~C.~B., \& {Nelson}, R.~P. 2001, \aap,
  374, 1092

\bibitem[{{Tamayo} {et~al.}(2016){Tamayo}, {Silburt}, {Valencia}, {Menou},
  {Ali-Dib}, {Petrovich}, {Huang}, {Rein}, {van Laerhoven}, {Paradise},
  {Obertas}, \& {Murray}}]{2016ApJ...832L..22T}
{Tamayo}, D., {Silburt}, A., {Valencia}, D., {et~al.} 2016, \apjl, 832, L22

\bibitem[{{Ter Braak}(2006)}]{TerBraak2006}
{Ter Braak}, C. J.~F. 2006, Stat. Comput., 16, 239

\bibitem[{{Thommes} \& {Lissauer}(2003)}]{2003ApJ...597..566T}
{Thommes}, E.~W., \& {Lissauer}, J.~J. 2003, \apj, 597, 566

\bibitem[{{Veras}(2007)}]{2007CeMDA..99..197V}
{Veras}, D. 2007, Celestial Mechanics and Dynamical Astronomy, 99, 197

\bibitem[{{Vogt} {et~al.}(1994){Vogt}, {Allen}, {Bigelow}, {Bresee}, {Brown},
  {Cantrall}, {Conrad}, {Couture}, {Delaney}, {Epps}, {Hilyard}, {Hilyard},
  {Horn}, {Jern}, {Kanto}, {Keane}, {Kibrick}, {Lewis}, {Osborne},
  {Pardeilhan}, {Pfister}, {Ricketts}, {Robinson}, {Stover}, {Tucker}, {Ward},
  \& {Wei}}]{1994SPIE.2198..362V}
{Vogt}, S.~S., {Allen}, S.~L., {Bigelow}, B.~C., {et~al.} 1994, in \procspie,
  Vol. 2198, Instrumentation in Astronomy VIII, ed. D.~L. {Crawford} \& E.~R.
  {Craine}, 362

\bibitem[{Vrugt {et~al.}(2009)Vrugt, ter Braak, Diks, Robinson, Hyman, \&
  Higdon}]{VrugtEtAl:2009}
Vrugt, J.~A., ter Braak, C. J.~F., Diks, C. G.~H., {et~al.} 2009, International
  Journal of Nonlinear Sciences and Numerical Simulation, 10, 273

\bibitem[{{Wisdom} \& {Holman}(1991)}]{1991AJ....102.1528W}
{Wisdom}, J., \& {Holman}, M. 1991, \aj, 102, 1528

\end{thebibliography}

\end{document}